\documentclass{aa}
\usepackage{latexsym,subfigure,epsfig,graphics}
\usepackage{graphicx}
\usepackage{natbib}
\bibpunct{(}{)}{;}{a}{}{,}

\newcommand{\Msun}{\>{\rm M_{\odot}}}
\newcommand{\kms}{\>{\rm km}\,{\rm s}^{-1}}

\begin{document}
   \title{HST observations of nuclear stellar disks}

%   \subtitle{}

   \author{Davor Krajnovi\'c\inst{1}
          \and
          Walter Jaffe\inst{1}
          }

   \offprints{D. Krajnovi\'c,\\
   \email{davor@strw.leidenuniv.nl}}

   \institute{$^1$Sterrewacht Leiden, Postbus 9513, 2300 RA Leiden, The Netherlands\\
             }
\date{Received 2004 March 1 ;  Accepted 2004 August 17}

\abstract{ 
We present observations of four nearby early-type galaxies with
previously known nuclear stellar disks using two instruments on-board
the Hubble Space Telescope. We observed NGC~4128, NGC~4612, and
NGC~5308 with the Wide Field Planetary Camera 2, and the same three
galaxies, plus NGC~4570, with the Space Telescope Imaging
Spectrograph. We have detected a red nucleus in NGC~4128, a blue
nucleus in NGC~4621, and a blue disk in NGC~5308. Additionally, we
have discovered a blue disk-like feature with position angle
$\sim15\degr$ from the major axis in NGC~4621. In NGC~5308 there is
evidence for a blue region along the minor axis. We discovered a blue
transient on the images of NGC~4128 at position $0\farcs14$ west and
$0\farcs32$ north from the nucleus. The extracted kinematic profiles
belong to two groups: fast (NGC~4570 and NGC 5308) and kinematically
disturbed rotators (NGC~4128 and NGC~4621). We report the discovery of
a kinematically decoupled core in NGC~4128. Galaxies have mostly old
(10-14 Gyr) stellar populations with large spread in metallicities
(sub- to super-solar). We discuss the possible formation scenarios,
including bar-driven secular evolution and the influence of mergers,
which can explain the observed color and kinematic features.
   
 }

\titlerunning{}
\authorrunning{Krajnovi\'c \& Jaffe}

\maketitle
%
%________________________________________________________________

\section{Introduction}
\label{s:intro}
The decade of Hubble Space Telescope (HST)
observations have revealed the existence of small scale nuclear
stellar disks in early-type galaxies. This discovery was an important
step in the long process of recognizing the complexity of early-type
galaxies. Ground-based studies preceding the HST era, having lower
resolution and polluted by typically $>$ 1\arcsec\, seeing, already
recognized two distinct classes of elliptical galaxies
\citep{1983ApJ...266...41D} that differed in photometric appearance --
disky vs. boxy -- and kinematic properties -- rotationally
vs. pressure supported -- \citep{1988A&A...193L...7B,
1989A&A...217...35B}. Follow-up studies discovered the existence of
embedded stellar disks in elliptical and lenticular galaxies
\citep{1995A&A...293...20S,1996A&A...310...75S}. These disks, although
similar to their counterparts in spiral and S0 galaxies, have smaller
scale length and higher central surface brightness. They often do not
follow the same exponential profiles, and are closer to $r^{1/4}$
profiles, reflecting formation in a different potential: dark halo
potentials for disks in late-types, and bulge potentials for disks in
early-type galaxies \citep{1995A&A...293...20S}. The existence of the
embedded disks also supports the idea of the morphological connection
between spiral, lenticular and elliptical galaxies
\citep{1996ApJ...464L.119K}.

The properties of the nuclear regions (inner few 100 pc) of early-type
galaxies are, however, not easily accessible from the ground. High
resolution imaging surveys with HST discovered small scale nuclear
stellar disks in early type galaxies (\citealt{1994AJ....108.1567J};
\citealt{1994AJ....108.1579V}; \citealt{1995AJ....110.2622L};
\citealt{2001AJ....121.2431R},hereafter R01). They were followed by
detailed photometric and kinematic studies on a few individual
objects, principally NGC~4342 \citep[][hereafter
BJM98]{1998MNRAS.300..469S, 1998MNRAS.293..343V}, NGC~4570
\citep[BJM98\nocite{1998MNRAS.293..343V};][]{1998MNRAS.300..469S,
1998MNRAS.298..267V}, NGC~4594 \citep{1986AJ.....91..777B,
1988ApJ...335...40K, 1996A&A...312..777E}, NGC~7332
\citep{1994AJ....107..160F, 2004MNRAS.350...35F}. A detailed study of
early-type galaxies with kinematicaly distinct components
\citep{1997ApJ...491..545C, 1997ApJ...481..710C} found photometric
evidences for faint nuclear stellar disks in a number of dust free
galaxies.

The next step was a search for embedded nuclear stellar disks in
bulges of spiral galaxies. The high resolution studies of spiral
galaxies with HST showed that a significant fraction of galaxies
classified as early-type spirals have a rich variety of central
properties, and show little evidence for r$^{1/4}$ law expected for
smooth bulges \citep{1998AJ....116...68C}. Similarly,
\citet{2003ApJ...582L..79B} found moderately large fraction (34\%) of
nuclear bars or disks in their HST near-infrared survey of S0 -- Sbc
galaxies. Using archival HST imaging, \citet{2002ApJ...573..131P}
reported evidence for nuclear disks in three early-type spirals and
concluded that the disks are restricted to S0 and unbarred spiral
galaxies. Having in mind that nuclear stellar disks are detectable
only when seen nearly edge on \citep{1990ApJ...362...52R}, they appear
to be very common, perhaps universal, in flattened ellipticals and S0s.

The large fraction of detected nuclear disks in early-type galaxies
(51\% in the R01\nocite{2001AJ....121.2431R} sample) presents the
questions: how and when did the nuclear stellar disks form?  Nuclear
stellar disks are found in S0 and disky ellipticals, but they are not
simple extensions of large scale disks to the center of the
galaxy. Often there is clear photometric and kinematic evidences for
double disk structures \citep{1994AJ....108.1579V,
1995A&A...293...20S, 1998MNRAS.300..469S}, where the double disk
structures are represented by two morphologically separated disks,
having different scale lengths, lying in nearly same plane, but
possibly with different inclinations, and having an inner/outer
separation radius between the disks. On the other hand,
\citet{2002AJ....124...65E} show evidences that some inner disks seen
in edge-on galaxies could be bars mistaken for disks. In any case, the
two dynamically different structures are not easily distinguished in
all cases.

Inner disks are also found inside bars or rings
\citep{1998MNRAS.298..267V, 1999ApJ...521L..37E,
2003ApJ...597..929E}. This is important for understanding their
formation. Nuclear stellar disks could be the result of mergers in
hierarchical galaxy formation scenario: accretion of gas during the
merger which settles in the principal plane of the galaxy and then
makes stars. On the other hand, disks could be formed from the galaxy
material transported to the nucleus by a bar, or perhaps from a
mixture of these processes, in which a bar fuels the center
effectively with gas captured at some previous epoch. Whatever
scenario we choose, it has to be consistent with the high
metallicities seen in the disks
\citep[BJM98\nocite{1998MNRAS.293..343V};][]{1996A&A...312..777E} as
well as their blue colors (BJM98\nocite{1998MNRAS.293..343V}; Kormendy
et al. 2002) implying younger stellar populations.

Nuclear stellar disks with their cold dynamical properties and high
surface brightness provide an excellent measure of the central
mass-to-light ratio, as well as of the mass of the central black hole
\citep{1996MNRAS.283..381V}. A few studies used this to determine the
mass of the back holes in galaxies with nuclear stellar disks
\citep{1996ApJ...473L..91K, 1996ApJ...459L..57K, 1999ApJ...514..704C,
1999MNRAS.303..495E}.

%%% Table 1.%%%%%%%%%%%%%%%%%%%%%%%%%%%%%%%%%%%%%%%%%%%%%%%%%%%%%%%%%%%%%

\begin{table}

 \caption[]{The properties of sample galaxies }
  \label{t:prop}
$$
  \begin{array}{cccccccc}
    \hline
    \hline
    \noalign{\smallskip}
    $galaxy$ & $type$ & $M$_{B} & $B-V$ & $v$_{rad} & $PA$ & $D$ &$scale$\\
    (1)&(2)&(3)&(4)&(5)&(6)&(7)&(8)\\
    \noalign{\smallskip}
    \hline
   \hline
     \noalign{\smallskip}
    $NGC~4128$ & S0   & -19.89 & 1.02 & 2610 &  58 & 36.3 & 175.7\\
    $NGC~4570$ & S0   & -20.39 & 0.97 & 1811 & 159 & 25.2 & 121.9\\
    $NGC~4621$ & E    & -20.49 & 0.97 &  524 & 165 &  7.3 &  35.3\\
    $NGC~5308$ & E-S0 & -20.38 & 0.93 & 2299 &  60 & 31.9 & 154.8\\

    \noalign{\smallskip}
    \hline
   \hline
   \end{array}
$$

{Notes -- Col.~(1): galaxy name; Col.~(2): morphological type;
Col.~(3): absolute B-magnitude; Col.~(4): apparent B-V color within
the effective aperture in which half of the B-flux is emitted;
Col.~(5): radial velocity (cz) in km s$^{-1}$ corrected for LG infall
onto Virgo; Col.~(6): major axis position angle in degrees;
Col.~(7): distance in Mpc, as derived from radial velocity (column 5)
using Hubble constant H$_{0}$=72 km s$^{-1}$ Mpc$^{-1}$
\citep{2001ApJ...553...47F}; Col.~(8): distance scale in pc
arcsec$^{-1}$. Values listed in columns 2 -- 6 are taken from
Lyon/Meudon Extragalactic Database (LEDA).}
\end{table}
%%%%%%%%%%%%%%%%%%%%%%%%%%%%%%%%%%%%%%%%%%%%%%%%%%%%%%%%%%%%%%%%%%%%%%%%%

Early-type galaxies are also interesting for studying stellar
populations. The absence of strong and continuous star-formation as
well as emission line gas makes it easier to investigate the formation
history and the connection between the photometric morphology,
dynamical structures and corresponding stellar populations.

In order to increase the available dataset and to investigate the
dependencies between the kinematics and line-strengths, as well as to
determine the mass of black holes, we obtained high resolution spectra
of four galaxies known to have nuclear stellar disks from the
R01\nocite{2001AJ....121.2431R} sample: NGC~4128, NGC~4570, NGC~4621
and NGC~5308. In addition, we also imaged three of the galaxies,
except NGC~4570 which was thoroughly investigated in the previous
studies \citep[BJM98,][]{1998MNRAS.300..469S, 1998MNRAS.298..267V}. In
this paper we present data observed with two instruments on-board HST
during Cycle 9 (Program ID 8667) and concentrate on the photometric
and spectroscopic properties and dependencies. Dynamical modeling of
the galaxies with the purpose of determining the masses of the central
black holes will follow in a separate paper.

Section~\ref{s:wfpc2} presents broad band photometry, data reduction,
isophotal analysis and color images. Section~\ref{s:stis} deals with
spectroscopic observations, data reduction, extraction of kinematics
and measurements of line-strengths. Section~\ref{s:discus} presents a
discussion of the results for individual galaxies. The conclusions and
summary of the work are presented in Section~\ref{s:con3}.

\section{WFPC2 broad band imaging}
\label{s:wfpc2}
The Wide Field Planetary Camera 2 (WFPC2) observations (Biretta et
al. 1996) included imaging in V (F555W) and B band (F450W). The
general properties of the sample galaxies are presented in
Table~\ref{t:prop}, the observations are summarized in
Table~\ref{t:wfpc2} and the details of the filter properties are
listed in Table~\ref{t:filters}. The centers of the galaxies were
positioned on the PC CCD. The size of the PC CCD is $800\times800$
pixels of $0\farcs0455\times 0\farcs0455$. All exposures were taken
with the telescope in fine lock.  In addition to newly acquired data,
we used existing archival I (F814W) band images for NGC~4621 (Program
ID 8212, PI Ajhar) and NGC~5308 (Program ID 5512, PI Faber). There
were no I band observations for NGC~4128 in the archive.

\subsection{Data reduction}
\label{ss:reduc}
The images were reduced through the standard HST/WFPC2 pipeline. Upon
request of the data, the On-The-Fly reprocessing system re-reduced the
data using the best calibration files. The standard reduction steps
include correction for analog to digital conversion error, bias and
dark current subtraction and flat-fielding (for more description see
\citet{1995PASP..107.1065H}). Our observations were divided (CR-SPLIT)
in two (per filter). To combine them and to remove the cosmic rays we
used a set of IDL routines from The IDL Astronomy User's Library
\citep{1993adass...2..246L}\footnote{{\it
http://idlastro.gsfc.nasa.gov}}. The WFPC2 images (PC, WF2, WF3, and
WF4 for both CR-SPLIT sections) are cross-correlated to determine a
possible shift between the exposures, aligned and combined removing
the cosmic rays using an IDL equivalent of the IRAF task CRREJ.

%%% Table 2.%%%%%%%%%%%%%%%%%%%%%%%%%%%%%%%%%%%%%%%%%%%%%%%%%%%%%%%%%%%%%

\begin{table}

 \caption[]{The summary of HST/WFPC2 observations }
  \label{t:wfpc2}
$$
  \begin{array}{cccccc}
    \hline
   \hline
     \noalign{\smallskip}
    $galaxy$ & $filter$ & $date$ & $time [s]$ & $ \# exp$\\
    \noalign{\smallskip}
    \hline
   \hline
     \noalign{\smallskip}
    $NGC~4128$ & F450W & 14.05.2001 & 1400 & 2\\
               & F555W & 14.05.2001 & 800  & 2\\
    $NGC~4621$ & F450W & 14.05.2001 & 1200 & 2\\
               & F555W & 14.05.2001 & 800  & 2\\
    $NGC~5308$ & F450W & 07.05.2001 & 1400 & 2\\
               & F555W & 07.05.2001 & 800  & 2\\

    \noalign{\smallskip}
    \hline
   \hline
   \end{array}
$$

\end{table}
%%%%%%%%%%%%%%%%%%%%%%%%%%%%%%%%%%%%%%%%%%%%%%%%%%%%%%%%%%%%%%%%%%%%%%%%%

We then constructed color images: B-I, V-I and B-V. To construct
the color images we had to align the individual images very
precisely. This was achieved by rotating the original images for the
difference in the telescope orientation angle, sub-sampling pixels by
a factor of six and cross-correlating images to find the shift. After
all shifts were applied, the images were rebinned to the original
pixel size. Both images used for the construction of a color image
were initially convolved with the PSF of the other image. The PSFs
were constructed using Tiny Tim software (Krist \& Hook 2001). The raw
counts of the images were converted into Johnson-Cousins B,V and I
magnitudes following the guidelines given by Holtzman et al. (1995),
using the zero points as given by
\citet{2000PASP..112.1397D}\footnote{{\it
http://www.noao.edu/staff/dolphin/wfpc2\_calib/}} and iterating the
calibration until convergence. Note that the iteration was not
performed for the B-I color images, because there were no published
transformation for the F450W filter using B and I. We estimate our
relative photometric accuracy to be $\approx 0.02$ mag, while the
absolute uncertainty is $\approx 0.05$ mag (and $\approx 0.1$ for B-I
color). In Figs.~\ref{f:isophote} and \ref{f:nuc_disk} we present
WFPC2 observations, isophotal analysis, broadband and color images of
three observed galaxies.

\subsection{Isophotal analysis}
\label{ss:iso}
\begin{figure*}[!t]
\begin{center}
   \psfig{file=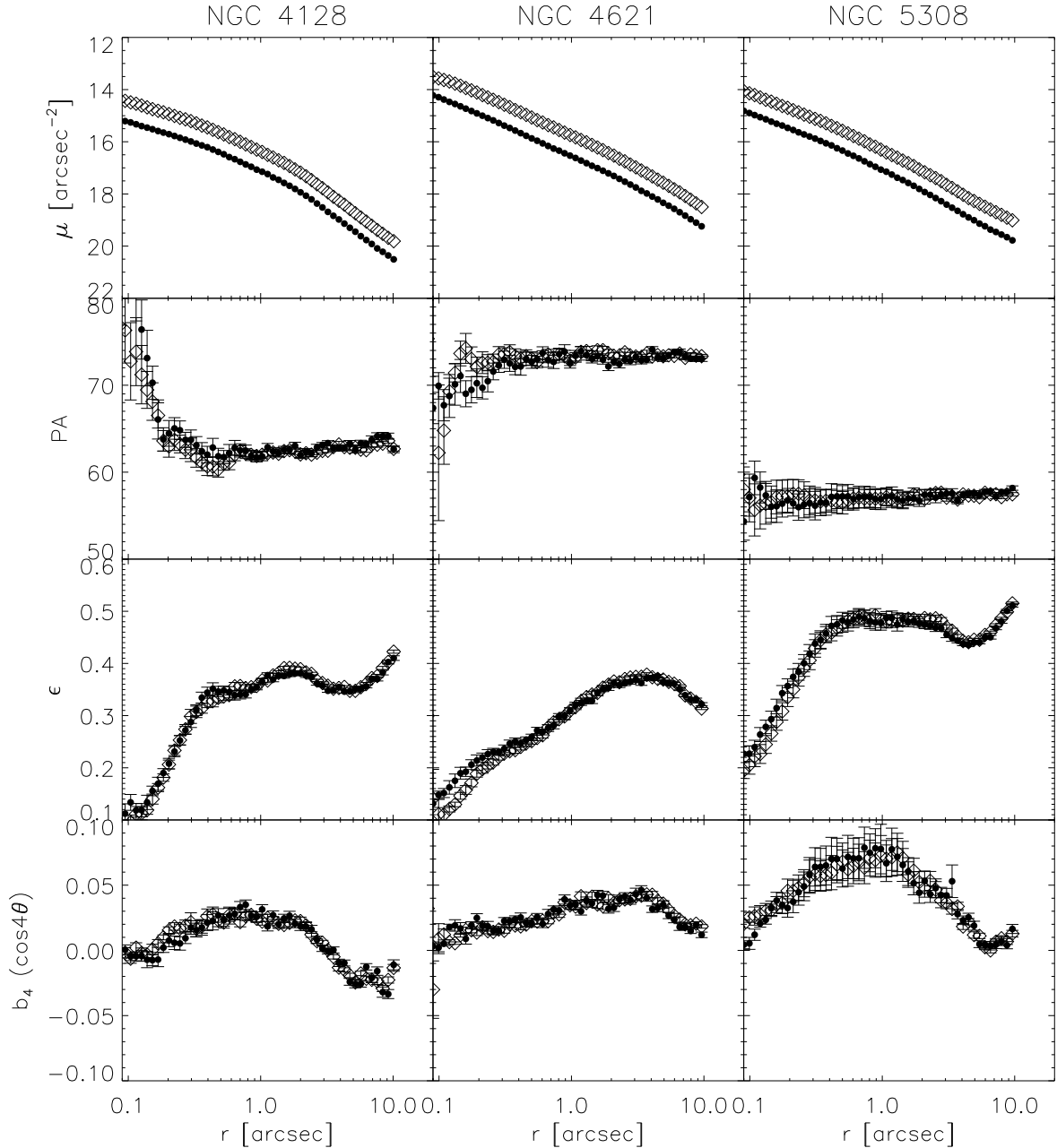, width=18cm}%, width=\textwidth}
   \caption{\label{f:isophote} Isophotal analysis results for
   NGC~4128, NGC~4621 and NGC~5308 in two observed filters: B (F450W)
   and V (F555W). First row: surface brightness profiles. Filled
   symbols correspond to B and open symbols to V filter. Errors are
   smaller then the symbols. Second row: position angles measured east
   of north; in the case of NGC~4621 we added 90\degr\, to the
   measured position angle (east of north) for presentation
   purposes. Third row: ellipticity. Fourth row: higher order
   parameters $b_{4}$ (the coefficient of $cos(4\theta)$), showing
   deviations from perfect ellipses.}
\end{center}
\end{figure*}

In order to investigate the disky structure of the galaxies we used
the IRAF task `ellipse' to perform isophotal fits to the light
distributions. We measured the ellipticity and position angle of the
isophotes, as a function of radius. The method (for a full description
see \citet{1987MNRAS.226..747J}) first fits elliptical isophotes to a
Fourier expansion of first and second order terms. The next step in
the method is to measure the higher order terms of the Fourier
expansion. The pure ellipse is given by the first two order terms in
the expansion. Any non zero values of the higher order ($>2$) terms
means a deviation from the perfect
ellipse. \citet{1990AJ....100.1091P} and \citet{1994A&AS..105..341G}
found that $cos3\theta$ terms ($b_{3}$) are sensitive to the presence
of dust (as well as the difference between the higher order terms in
different bands), while the $cos4\theta$ terms ($b_{4}$) describe the
shape by distinguishing boxy ($b_{4} < 0$) from disky ($b_{4} > 0$)
galaxies \citep[e.g.][]{1985MNRAS.216..429L,1988A&A...193L...7B}. The
isophotal parameters for three galaxies are shown in
Fig~\ref{f:isophote}. Different studies
\citep[][R01]{1994AJ....108.1579V} showed that although the isophote
parameters can be fitted down to 0\farcs03, they are not reliable and
only values $> 0\farcs2$ should be used for analysis.
%%% Table 3.%%%%%%%%%%%%%%%%%%%%%%%%%%%%%%%%%%%%%%%%%%%%%%%%%%%%%%%%%%%%%

\begin{table}

 \caption[]{The properties of the WFPC2 broad band filters used (from
 WFPC2 Instrument Handbook), showing the central wavelength and the
 size of the wavelength window covered by the given filters }
  \label{t:filters}
$$
  \begin{array}{cccc}
    \hline
   \hline
     \noalign{\smallskip}
    $filter$ &\lambda_{cen} [$\AA$] & \Delta \lambda $[\AA]$ & $band$\\
   \noalign{\smallskip}
    \hline
   \hline
     \noalign{\smallskip}
    $F450W$ & 4410 & 925 & B\\ 
    $F555W$ & 5202 & 1223& V\\
    \noalign{\smallskip}
    \hline
   \hline
   \end{array}
$$
\end{table}
%%%%%%%%%%%%%%%%%%%%%%%%%%%%%%%%%%%%%%%%%%%%%%%%%%%%%%%%%%%%%%%%%%%%%%%%%

%%%%%Figure 1.%%%%%%%%%%%%%%%%%%%%%%%%%%%%%%%%%%%%%%%%%%%%%%%%%%%%%%%%%
%%%%%%%%%%%%%%%%%%%%%%%%%%%%%%%%%%%%%%%%%%%%%%%%%%%%%%%%%%%%%%%%%%%%%%%%%%

The galaxies were selected on the basis of having a nuclear stellar
disk and our isopohotal analysis agrees well with the R01 results. For
all three galaxies the $b_{3}$ terms are consistent with being zero in
the reliable range ($> 0\farcs2$). The $b_{4}$ terms have clear disky
deviations, but there are differences between the galaxies. NGC~4128
shows the smallest positive values of $b_{4}$ coefficient, and is the
only galaxy where $b_{4}$ becomes negative. It drops below zero at a
radius of $\sim 3\arcsec$ from the center. The isophotes remain boxy
in the rest of the investigated range (3 -- 10\arcsec). The fourth
order term in NGC~4621 is positive in the investigated range, although
it starts to decrease beyond $3\arcsec$. NGC~5308 shows the sharpest
rise of the $b_{4}$ coefficient, but it drops to zero around
7\arcsec\, before it rises again. The analysis of photometric higher
order terms suggests the nuclear stellar disks are separated from the
large scale disks.

The coefficients of this isophotal analysis are one-dimensional
representations of 2D structures in the galaxies. In order to see more
clearly the nuclear disks we constructed residual B and V images by
subtracting a model galaxy from the original.  The model galaxy was
constructed using the IRAF task `bmodel'. It had the same luminosity
profile as the original galaxy, but it was constructed from perfect
ellipses (first two terms in the Fourier expansion). The contour maps,
shown in the middle row in Fig.~\ref{f:nuc_disk}, reveal the disky
structure that is responsible for the existence of the higher order
terms. These disky structures do not necessarily show the full disks,
because some information on the disks is also contained in the
ellipticity, which was subtracted by the model. However, to the first
approximation, these structures are a good representation of the
relatively faint nuclear stellar disks in the observed galaxies. The
disks have various sizes between 2 -- $5\arcsec$, corresponding to 150
-- 600 pc, and are in a good agreement with the spatial extent of
positive values of $b_{4}$ terms.  The noise structure in
Fig.~\ref{f:nuc_disk} perpendicular to the disks (along the minor
axis) is not real and is an artifact of subtracting a perfect
elliptical structure from a very disky one.

\subsection{Broad-band color images}
\label{ss:color}
\begin{figure*}[!t]
\begin{center}
   \psfig{file=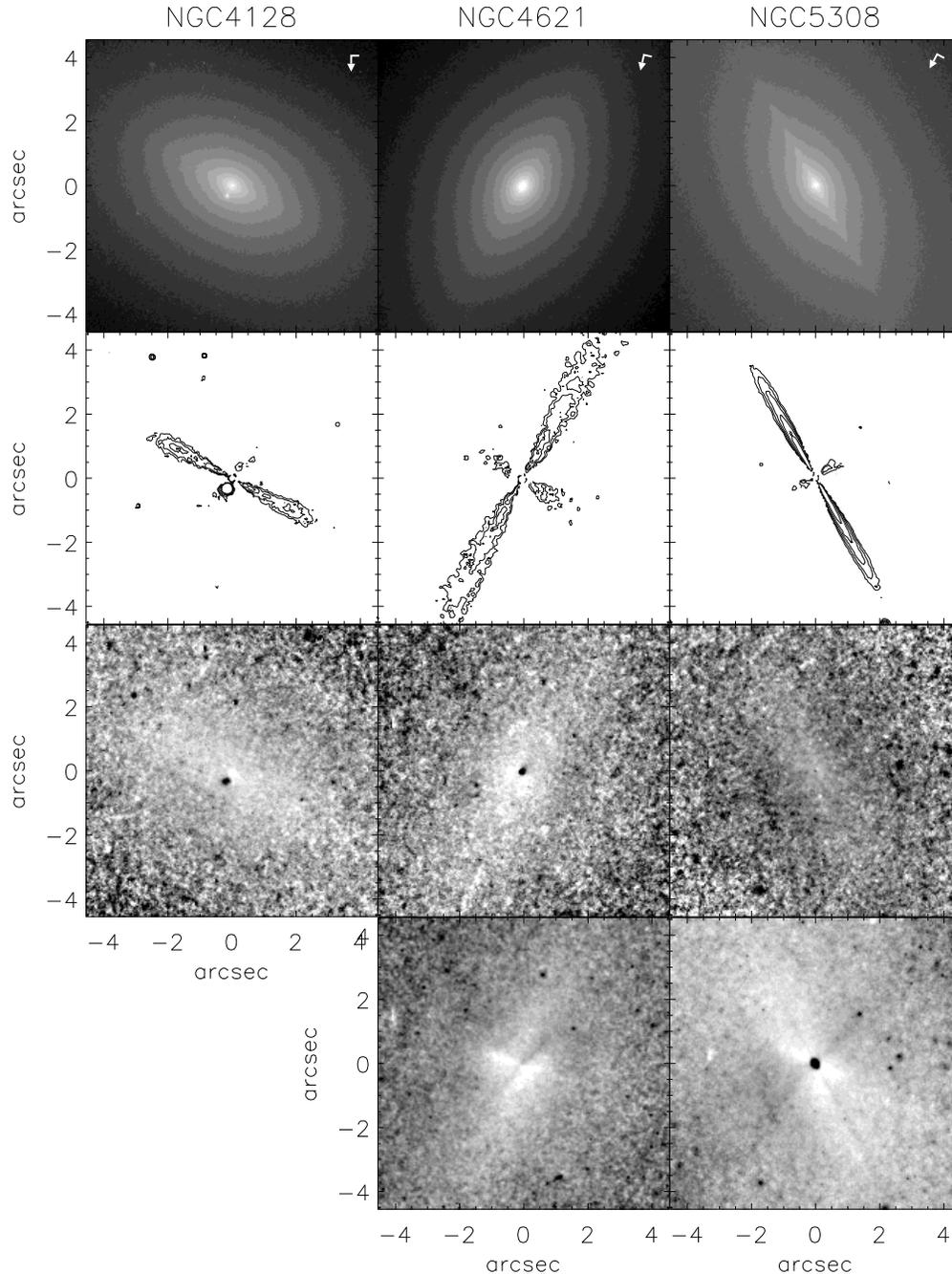, width=17.5cm}
   %,width=\textwidth}
   \caption{\label{f:nuc_disk} WFPC2 observation of NGC~4128, NGC~4621,
   NGC~5308. First row: images in the V (F555W) filter. The arrows and
   their associated dashes mark the North and East orientation of the
   images. Second row: residual stellar disks after subtraction of
   perfect elliptical model galaxy (in V band). Contours were slightly
   smoothed and logarithmically scaled with steps of 0.5
   magnitudes. Third row: B-V images, with grayscale such that
   brighter means redder. The blue feature next to the center of
   NGC~4128 is due to the transient and is discussed in Appendix
   B. The plotted range is (in magnitudes) 0.6 (black) to 1.0 (white)
   for NGC~4128, 0.5 (black) to 1.05 (white) for NGC~4621, and 0.5
   (black) to 0.98 (white) NGC~5308. Fourth row: B-I images. The
   central blue feature on the NGC~5308 color map is an artifact of
   saturation and was not considered in the analysis. Range is from
   1.93 (black) to 2.16 (white) for NGC~4621, and 1.1 (black) to 2.07
   (white) for NGC~5308.}
\end{center}
\end{figure*}

We constructed several broad band images using our data and archival
images as described in Section~\ref{ss:reduc}. In the bottom two rows
of Fig.~\ref{f:nuc_disk} we show B-V and B-I images for all three
galaxies except for NGC~4128 which does not have archival I band
images. These color images are color coded such that lighter shades
indicate redder colors.

%%%%%Figure 2%%%%%%%%%%%%%%%%%%%%%%%%%%%%%%%%%%%%%%%%%%%%%%%%%%%%%%%%%%%%%%
%%%%%%%%%%%%%%%%%%%%%%%%%%%%%%%%%%%%%%%%%%%%%%%%%%%%%%%%%%%%%%%%%%%%%%%%%

\subsubsection{B-V color images}
\label{sss:b-v}
The nuclei of the galaxies show different structures in the B-V
color. The nucleus of NGC~4128 is redder then the surrounding bulge,
NGC~4621 has a convincingly blue nucleus, while NGC~5308 is rather
uniform with a slightly bluer central pixel. The magnitudes measured
on the constructed color images are presented in
Table~\ref{t:color}. We measured the color within a small circular
aperture (6 pixels in diameter) to point out the subtle differences in
colors between different regions and features. The blue feature on the
color image of NGC~4128 corresponds to a transient object detected in
the galaxy. The nature of the transient is not known (see Appendix B
for a detailed discussion). In NGC~4128 isochromes trace the nuclear
disk, while in the case of NGC~5308 there is weak evidence of a blue
disk outside 0\farcs3 of the nucleus. The difference is $\approx0.02$
magnitudes. NGC~4621 is a special case different from both previous
galaxies. Here we have a prominent blue nucleus on top of the red
bulge. The other particularity of this galaxy is in the blue component
that stretches southwards from the nucleus. The average difference
between this component and the rest of the bulge is again
$\approx0.02$.

\subsubsection{B,V-I color images}
\label{sss:B-I}
A bigger color difference is seen in the B-I and V-I images which we
constructed for NGC~4621 and NGC~5308. The archival I-band images of
NGC~4621 were of good quality, while images of NGC~5308 were
saturated, and the very central parts of the images are not reliable.

The interesting blue feature southwards from the nucleus on the B-I
image of NGC 4621 is more prominent and we can quantify its
extent. The color difference between the red bulge-like background and
the blue patch is here $\approx0.12$ mag in its extreme. The extent of
the feature is $\approx1\farcs35$ on each side of the nucleus, being
somewhat bluer on the south side. The feature lies generally in the
north-south direction making an angle of 15\degr\, to the major
axis. It is in no obvious plane of symmetry of the galaxy. The rest of
the bulge is red, but a region of enhanced red color lies in the
east-west direction. The angle between this reddest region and the
major axis of about 103\degr, making the angle between the bluest and
the reddest regions about 118\degr. The V-I color image confirms this
finding, also noticed by \citet{2002A&A...396...73W}, hereafter
WEC02\nocite{2002A&A...396...73W}, on their V-I images. We postpone
the discussion and interpretation of these features to
Section~\ref{ss:ngc4621}.

NGC~5308 is an equally interesting case. The center of the I band
image was saturated and the blue nucleus is an artifact. Generally,
one has to be very careful in interpreting the saturated images. Aware
of problems in dealing with saturated images, we used them to verify
the faint suggestion from the B-V images of the blue disk. We do not
consider the central $0\farcs5$ of the B-I and V-I color images (the
area somewhat larger than the blue dip in the images), but concentrate
on the larger-scale features. 

Along the minor axis of the galaxy, the B-I color image also reveals
blue regions on each side of the nucleus, as well as a red feature in
the east-west direction. These features were not anticipated and we
checked whether they are real or artifacts of saturation since they
are close to the nucleus (r $<$ 1\arcsec). We first constructed color
image from unconvolved F4550W and F814W images. On the resulting color
image, there was a hint of the east-west red feature. The other
approach included deconvolution of the B and V images using
Richardson-Lucy algorithm with 20 iterations. The resulting images
were convolved with the PSF of the I image, and used to make color
images (B-I and V-I). On both color images, next to the red east-west
feature, we also detected the blue region on the minor axis. These
tests suggest the color features on the last panel of the
Fig.~\ref{f:nuc_disk} are real, although could be augmented by the
convolution process due to its proximity to the saturated nucleus.

%%% Table 4.%%%%%%%%%%%%%%%%%%%%%%%%%%%%%%%%%%%%%%%%%%%%%%%%%%%%%%%%%%
\begin{table}

 \caption[]{Color in magnitudes measured within apertures of 6 pixels
 ($0\farcs273$) in diameter}
  \label{t:color}
$$
  \begin{array}{c|cc|cc|cc}
    \hline
   \hline
     \noalign{\smallskip}

    \multicolumn{1}{c|}{galaxy} &\multicolumn{2}{c|}{$B - V$ } &
    \multicolumn{2}{c|}{$V - I$}& \multicolumn{2}{c}{$B - I$}\\

     & center & average & center & BF & bulge & BF \\ 
    (1)&(2)&(3)&(4)&(5)&(6)&(7)\\ 
    \noalign{\smallskip}
    \hline    \hline
 \noalign{\smallskip}
  
    $NGC~4128$ & 1.00 & 0.95 &-- &-- &-- &-- \\
    $NGC~4621$ & 0.87 & 0.99 & 1.22 & 1.16 & 2.14 & 2.05\\
    $NGC~5308$ & 0.97 & 0.94 &-- &-- & -- & -- \\

    \noalign{\smallskip}
    \hline
   \hline
   \end{array}
$$ 

{Notes -- Col.~(1): galaxy name; Col.~(2): B--V color measured at the
center of the galaxies; Col.~(3): B--V color averaged over 8 apertures
placed around center on a square grid centered on the nucleus with
size $2\times14$ pixels ($\sim1\farcs27$), except for NGC~4128 where
the blue feature next to the center was excluded; Col.~(4): V--I color
measured at the center of galaxy; Col.~(5): V--I color measured at the
blue feature $0\farcs90$ from the center; Col.~(6): B--I color
measured at bulge of the galaxy; Col.~(7): B--I color measured at the
blue feature $0\farcs90$ from the center.}

\end{table}
%%%%%%%%%%%%%%%%%%%%%%%%%%%%%%%%%%%%%%%%%%%%%%%%%%%%%%%%%%%%%%%%%%%%%%%%%

The remaining and real (clearly visible on all test images), thin blue
feature follows the major axis of the galaxy, along which there is
clear evidence for a very thin disk on all scales, from the nucleus
outwards. The position angle of the thin blue component is the same as
of the nuclear stellar disk.

The marginal difference in color ($\approx0.01-0.02$ mag) is almost
razor sharp and it looks like the signature of the thin disk visible
also in the residual image (second row of Fig.~\ref{f:nuc_disk}). We
compared the sizes of the disk in the residual image and the blue
feature in the B-I color image. The comparison is made by extracting
and averaging together several profiles of intensity and color
perpendicular to the disk, on both sides of the nucleus (avoiding the
central 2\arcsec). The final profiles were fitted with Gaussians. The
size (FWHM) of the disk feature on the residual disk image is
$\sim5.6$ pixels and the size of the color feature is $\sim4.4$
pixels. These numbers correspond to 0\farcs25 and 0\farcs20
respectively, and they are in good agreement, enforcing the connection
between the components. At the distance of NGC~5308, the blue
component in the color image is approximately 30 pc thick.

\begin{figure}
  \centering
   \psfig{file=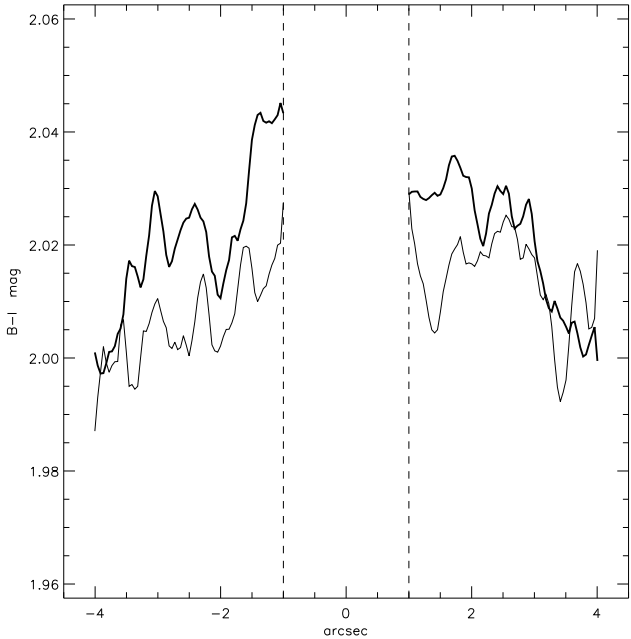, width=\columnwidth}
   \caption{\label{f:ngc5308_slit} Comparison of B-I color profiles
   extracted (and smoothed) along the disk (thin line) and parallel to
   disk (thick line) in NGC~5308. The slit-like apertures were 200x1
   pixels in size. One slit (thin line) was placed along the major
   axis (and the disk), while two other slits (averaged together and
   presented by the thick line) were placed parallel to the major
   axis, 10 pixels (0\farcs45) above and below it. Vertical dashed
   lines show the nuclear region excluded from the measurement due to
   the saturation effects.\looseness=-2}
\end{figure}

A way to quantify the relative difference in color between the disk
and the bulge is shown on Fig.~\ref{f:ngc5308_slit}. We measured the
color along slit-like apertures (1 pixel wide) along the disk and
parallel to it, on both sides of the nucleus. The central 1\arcsec\,
of all slits were omitted, and the two slits, positioned on each side
of the central slit, were averaged and presented as one color
profile. The disk is clearly bluer then the bulge in the inner
8\arcsec. Beyond 3\farcs5 on each side of the center colors of the
disk and bulge become similar.

%%%%%%Figure 3%%%%%%%%%%%%%%%%%%%%%%%%%%%%%%%%%%%%%%%%%%%%%%%%%%%%%%%%%%%%%%
%%%%%%%%%%%%%%%%%%%%%%%%%%%%%%%%%%%%%%%%%%%%%%%%%%%%%%%%%%%%%%%%%%%%%%%%%%5

Briefly summarizing, we list below the important observed color
features to which we refer later in the text:

\noindent {\bf NGC~4128} has (i) a red nucleus and (ii) a blue feature
0\farcs14 west and 0\farcs32 north of the galaxy center;

\noindent {\bf NGC~4621} has (i) a blue nucleus, (ii) an extended blue
component with PA $\approx150\degr$ (east of north), and (iii) an
extended red component with PA $\approx268\degr$ (east of north);

\noindent {\bf NGC~5308} has (i) an extended blue component along the
major axis and (ii) a wide blue component along the minor axis, (iii)
an extended red component with an east-west orientation.

\section{STIS spectroscopy}
\label{s:stis}
Spectra of all four galaxies were obtained using the Space Telescope
Imaging Spectrograph (STIS) with aperture 52\arcsec x0\farcs2 and
grating G430M (for details see \citet{1998ApJ...492L..83K}). The
observations are summarized in Table~\ref{t:stis_obs}. The
configuration of STIS and the properties of the grating are listed in
Table~\ref{t:stis_config}. With this setup we chose to observe Mg
lines at 5180 \AA\, rather then Ca lines at 8700 \AA\, which are
commonly used for extracting stellar kinematics. The reasons were:
(i) the Mg lines provide simultaneous kinematic data and a commonly
used index of stellar metallicity, (ii) STIS shows problems with
scattered light inside the CCD chip in the near infra-red which generates
artificial ``wings'' on spectral features, and (iii) the spatial
resolution along the slit in V-band is better by almost a factor of
two due to the decreased Airy disk size.

%%% Table 5.%%%%%%%%%%%%%%%%%%%%%%%%%%%%%%%%%%%%%%%%%%%%%%%%%%%%%%%%%
\begin{table}

 \caption[]{Summary of HST/STIS observations}
  \label{t:stis_obs}
$$
  \begin{array}{cccccrr}
    \hline 
   \hline
     \noalign{\smallskip}
    $galaxy$ & $slit$ & $date$ & $time$ & $\# exp$ & \Delta_{cen} & PA \\
    (1)&(2)&(3)&(4)&(5)&(6)&(7)\\
    \noalign{\smallskip}
    \hline
    \hline
    \noalign{\smallskip}
    $NGC~4128$ & cen & 02.12.2001 & 2568.5 & 3+3 &  0.00 & -112.9\\
               & pos & 02.12.2001 & 2700   & 3   &  0.16 & -112.9\\
               & neg & 02.12.2001 & 2697   & 3   & -0.40 & -112.9\\
    $NGC~4570$ & cen & 01.04.2001 & 2249.5 & 3+3 & -0.20 &  152.1\\
               & pos & 01.04.2001 & 2369   & 2   &  0.68 &  152.1\\
               & neg & 01.04.2001 & 2520   & 2   & -0.56 &  152.1\\
    $NGC~4621$ & cen & 01.04.2001 & 2260.5 & 3+3 & -0.12 &  161.1\\
               & pos & 01.04.2001 & 2380   & 3   &  0.40 &  161.1\\
               & neg & 01.04.2001 & 2520   & 3   & -0.44 &  161.1\\
    $NGC~5308$ & cen & 12.07.2000 & 2525.0 & 3+3 &  0.00 &   60.4\\
               & pos & 12.07.2000 & 2667   & 3   &  0.36 &   60.4\\
               & neg & 12.07.2000 & 2670   & 3   & -0.36 &   60.4\\

    \noalign{\smallskip}
    \hline
    \hline
  \end{array}
$$ 

{Notes -- Col.~(1): galaxy name; Col.~(2): the position of slit, cen
    -- center, pos -- positive offset, neg -- negative offset (see
    text for details); Col.~(3): date of observations; Col.~(4): total
    exposure time in seconds (exposure time of central slits is the
    average time of all added observations); Col.~(5): number of used
    observations; Col.~(6): distance of the center of the slit from
    the galaxy nucleus in arcsec; Col~(7): position angle of the slit
    in degrees east from north. }

\end{table}
%%%%%%%%%%%%%%%%%%%%%%%%%%%%%%%%%%%%%%%%%%%%%%%%%%%%%%%%%%%%%%%%%%%%%%%%%

\subsection{Data reduction}
\label{ss:redustis}
The galaxies were observed in a similar manner with a total of four
orbits per galaxy. Spectra were taken at three parallel positions per
galaxy during the four orbits. Each orbit was divided (CR-SPLIT) into
3 shorter exposures. Two orbits were used for the slit placed on the
center of the galaxies ({\it cen}), along the major axis. Between
orbits, the galaxy was shifted along the slit for about 0\farcs2, or 4
pixels, to get a better estimate of detector sensitivity variations
and to identify hot pixels. This strategy was not successful for the
case of NGC~4128, where the measured shift was $\sim1$\,pixel. The
remaining two orbits were split between two slit positions on either
side of the central slit, covering the bulge parallel to the nuclear
stellar disk.  One slit was targeted at the position +0\farcs3 away
from the central slit (positive offset -- {\it pos}), and the other at
the position -0\farcs3 away from the central slit (negative offset --
{\it neg}).

%%% Table 6.%%%%%%%%%%%%%%%%%%%%%%%%%%%%%%%%%%%%%%%%%%%%%%%%%%%%%%%%
\begin{table}

 \caption[]{The configuration of STIS and the properties of the
 grating  }
  \label{t:stis_config}
$$
  \begin{array}{lc}
    \hline
    \hline
    \noalign{\smallskip}
    $Quantity$ & $Value$\\
    \noalign{\smallskip}
    \hline
    \hline
    \noalign{\smallskip}
       $Aperture$                                      &52$x$0.2 \\
       $Grating$                                       &$G$430$M$ \\
       \lambda$-range [\AA]$                           &5050.4-5381.6\\
       \lambda_{cen}$ [\AA]$                           &5216\\
       $Scale$\, \Delta \lambda$ [\AA ~pixel$^{-1}$]$  &0.28 \\
       $Spatial scale [arcsec pixel$^{-1}$]$           &0.05\\
       $Comparison line FWHM [pixel]$                  &2.9 \\
       {\it R =} \lambda/\Delta \lambda                &6461\\
       $Instrumental dispersion [km s$^{-1}$]$         &19.76\\
    \noalign{\smallskip}
    \hline
   \hline
   \end{array}
$$

\end{table}
%%%%%%%%%%%%%%%%%%%%%%%%%%%%%%%%%%%%%%%%%%%%%%%%%%%%%%%%%%%%%%%%%%%%%%%%%

Most of the data reduction was performed by the HST/STIS calibration
pipeline CALSTIS (Hodge et al. 1998), including subtraction of
overscan, bias and dark, then flat-fielding, hot pixel and cosmic-ray
removal, absolute sensitivity calibration and wavelength
calibration. CR-SPLIT data sets were combined automatically in the
pipeline. The additional combination of the two central slit exposures
was performed manually outside the pipeline in IRAF, using the STSDAS
task MSCOMBINE. This task averages the exposures scaled by their
exposure times, and combines the separate exposures using a robust
sigma clipping rejection method. This was done on the files that have
\textit{\_crj} extension, i.e., after pipeline co-addition of CR-SPLIT
images and before the calibrations. The combined files were then
returned to CALSTIS for the calibrations. The same procedure, however,
was not possible on the side slits which were taken during only one
orbit. As there were still some cosmic rays left after the pipeline
reduction, we used Laplacian Cosmic Ray Identification (LAcosmic)
developed by \citet{2001PASP..113.1420V} to remove them. LAcosmic was
also applied on the \textit{\_crj} files. The detection limit for the
outliers was 3.5$\sigma$. To improve the quality of the images and
remove additional negative pixels we tried a few techniques. For
spectra with very low signal-to-noise it was possible to compare
different exposures (e.g. the two side slits) and to recognize the
same negative pixels and create a mask of them. For spectra with
higher signal to noise this was not effective and a different approach
was used. Using a boxcar filter we smoothed each LAcosmic-filtered
image. These images were subtracted from the corresponding original
LAcosmic-filtered images to emphasizes the outlying pixels. They were
flagged creating a mask image. Masked pixels were interpolated using
IRAF task FIXPIX. The resulting images were returned to CALSTIS and
processed to the end of the pipeline.

\begin{figure}
\centering
   \psfig{file=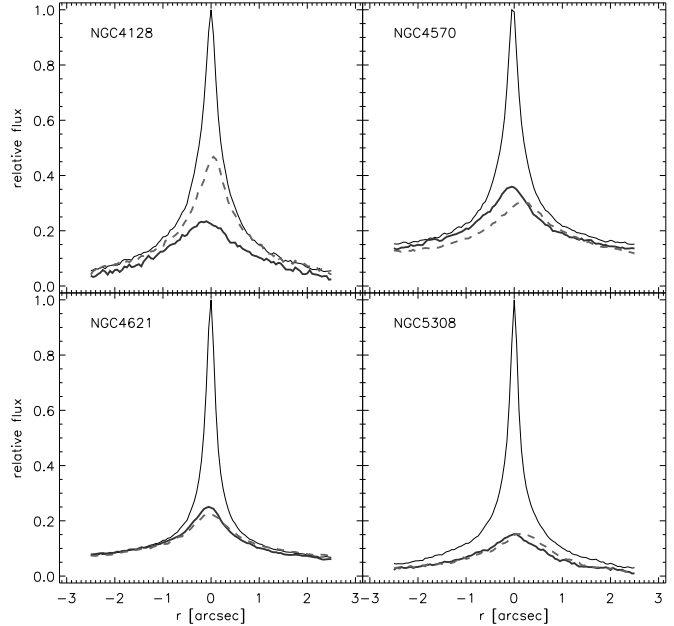, width=\columnwidth}
   \caption{\label{f:profs} Galaxy light profile along the STIS slit. The full
 spectral range was used. Thin dark line is the profile along the {\it
 cen} slit, thick gray lines are profiles along the side slits: dashed
 line along the {\it pos} slit and full line along the {\it neg} slit.
 The profiles were normalized to the maximum of the {\it cen} slit in
 order to emphasize the difference of intensities.}
\centering
\end{figure}

The final STIS light profiles are shown in Fig.~\ref{f:profs}. A
noticeable feature is the difference in the intensity of the side
slits. If the centering and shifting process worked properly, as the
light profiles of the galaxies are quite symmetric, it is expected
that the side slits should have very similar profiles. This is true
for the case of NGC~5308 and NGC~4621, suggesting those slits were on
similar but opposite positions. However, the other two galaxies show
significant deviations. It is therefore necessary to find the exact
positions of all slits.

%%%%%%%FIgure 4%%%%%%%%%%%%%%%%%%%%%%%%%%%%%%%%%%%%%%%%%%%%%%%%%%%%%%
%%%%%%%%%%%%%%%%%%%%%%%%%%%%%%%%%%%%%%%%%%%%%%%%%%%%%%%%%%%%%%%%%%%%%%%%%%%%%

\begin{figure}
\centering
   \psfig{file=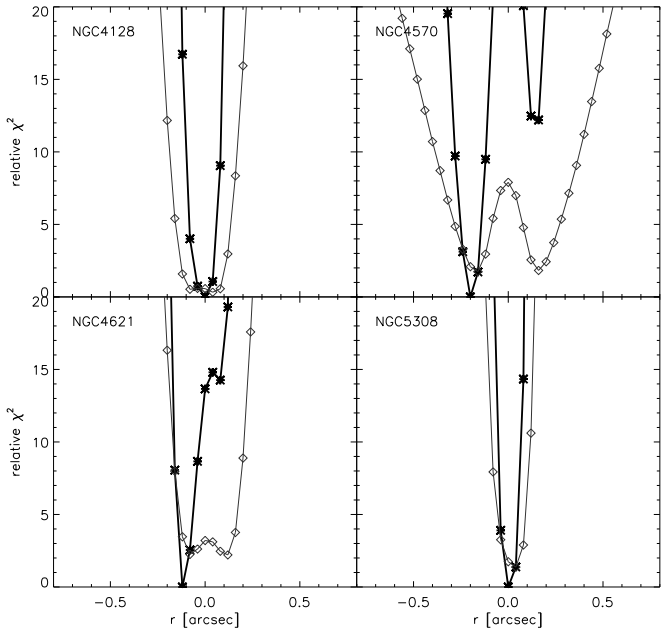, width=\columnwidth}
   \caption{\label{f:chi2}Plots of $\chi^{2}$ versus the positions of
   the slits for each galaxy. The line connecting the diamonds is the
   relative $\chi^{2}$ of the central slit. The line connecting
   asterisks is the total relative $\chi^{2}$ obtained by including
   the side slits in calculations.}
\centering
\end{figure}

We checked the actual positions of the slits by comparing the light
profiles from the STIS spectra with the WFPC2 images. The width of the
slit (dispersion direction) is 0\farcs2, which corresponds to 4 pixels
on the STIS CCD. We sub-sampled the image such that the slit width
projects to 5 pixels, in order to center the slits more
correctly. Summing up along the dispersion axis ($x$-axis on the CCD)
we created a STIS light profile. This was repeated for all slits. The
F555W images, which were used in the comparison, were accordingly
re-sampled. In each case we scanned the WFPC2 image by a combination
of the three slits, independently varying the distances between the
slits. The comparison of the STIS and WFPC2 profile was expressed by
the relative $\chi^{2}$ ({\it profile(WFPC2)/profile(STIS) - 1}). In
this process we assumed that the position angle of the telescope did
not change between different slit positions. The resulting $\chi^{2}$
estimates are shown in Fig.~\ref{f:chi2}. If the central slit is not
on the nucleus, the $\chi^{2}$ is expected to have double minima and
it is hard to distinguish which one is correct (NGC~4621 and NGC~4570
are clear examples). However, using the additional light profiles of
the side slits tightens the constraints producing one clear minimum,
which corresponds to the position of the central slit. The uncertainty
of our estimate is 0\farcs04. The positions of the central and the
side slits with respect to the nucleus are given in last two columns
of Table~\ref{t:stis_obs}. The slits in NGC~5308 and NGC~4128 are
centered on the galaxy nucleus, while for NGC~4621 and NGC~4570 the
slits were offset. The side slits were roughly on the requested
positions for NGC~4621 and NGC~5308. In the case of NGC~4128, the side
slits are the least symmetrically positioned and the {\it pos} slit is
almost coincident with the central slit; while in the case of the
NGC~4570 the positions of the side slits are the farthest apart, as
suggested from the profiles (Fig.~\ref{f:profs}).

%%%%%%%%Figure 5%%%%%%%%%%%%%%%%%%%%%%%%%%%%%%%%%%%%%%%%%%%%%%%%%%
%%%%%%%%%%%%%%%%%%%%%%%%%%%%%%%%%%%%%%%%%%%%%%%%%%%%%%%%%%%%%%%%%%%%%%

The general characteristic of the spectroscopic data is their low
signal-to-noise ratio ($S/N$). Only the major axis spectra have an
$S/N$ sufficient for the extraction of kinematics and line-strengths
as a function of radius. The side slits are much noisier and no
kinematic measurements were possible. It was possible however to
extract line-strength information from a few central rows, summed
together to increase $S/N$ creating one spectrum per side slit per
galaxy.

\subsection{Stellar kinematics}
\label{ss:kin}
All available information about the stellar kinematic properties of
galaxies are given by the line-of-sight velocity distribution
(LOSVD). The process of extracting kinematics is based on the
deconvolution of the observed galaxy spectra in order to recover the
full LOSVD. The idea behind this is that the galaxy spectrum can be
reproduced using a combination of several representative stellar
spectra convolved with the true LOSVD. Unfortunately, the LOSVD is not
{\it a priori} known and the deconvolution process is ill-determined,
being heavily dependent on the quality of the data. Over the last
thirty years a number of methods were invented to tackle the problem
and deliver the best possible estimates of the LOSVD (see
\citet{2003MNRAS.339..215D} for an overview of methods). Here we
choose to use a parametric method operating in the pixel space because
of the low $S/N$ of our data and very short wavelength range. We use
the penalized pixel fitting (pPXF) method
\citep{2004PASP..116..138C}. We derive the LOSVD parameterized by a
Gauss-Hermite series \citep{1993ApJ...407..525V,
1993MNRAS.265..213G}. The method finds the best fit to a galaxy
spectrum by convolving an optimal template spectrum with the
corresponding LOSVD given by the mean velocity $V$ and velocity
dispersion $\sigma$, as well as higher order Gauss-Hermite moments
$h_{3}$ and $h_{4}$. The higher order moments measure asymmetric and
symmetric deviation of the LOSVD from a Gaussian respectively.

An element which can heavily influence the extracted kinematics is the
stellar template used to convolve the LOSVD to reproduce the galaxy
spectra. There are methods, such as Fourier correlation quotient
\citep{1990A&A...229..441B} or Cross-correlation method
\citep{1995AJ....109.1371S}, which are less sensitive to template
mismatch. Pixel fitting techniques are much more sensitive to template
mismatch and it is crucial to have a good stellar template before
starting the extraction. The usual way is to observe a number of
representative stars (matching the spread in metallicity and age of
stars in the observed galaxy) with the same instrumental set-up and to
build an optimal template as a weighted linear combination of the
observed stellar spectra.

%%%%%%%%%%%FIgure 6%%%%%%%%%%%%%%%%%%%%%%%%%%%%%%%%%%%%%%%%%%
\begin{figure}
\centering
   \psfig{file=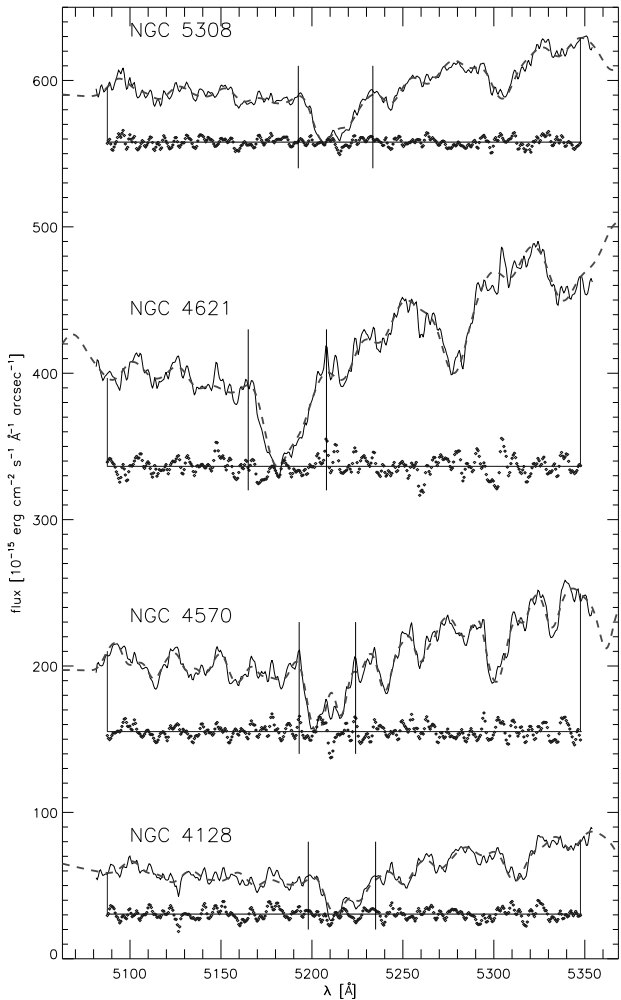, width=\columnwidth}
   \caption{\label{f:spec}Example of spectra for the four studied
   galaxies. From bottom up: NGC~4128, NGC~4570, NGC~4621 and
   NGC~5308. The spectra are shifted vertically to avoid overlap. The
   thin black lines indicate spectra of galaxies. The dashed thick
   lines are broadened optimal templates. The dots below the spectra
   are residuals of the fit (difference between galaxy spectra and
   optimal templates). Vertical and horizontal lines shows the region
   used in the fit. Two solid vertical lines crossing the spectra show
   the spectral regions excluded from the fit which tested the
   influence of Mgb region to the extracted kinematics. }
\centering
\end{figure}
%%%%%%%%%%%%%%%%%%%%%%%%%%%%%%%%%%%%%%%%%%%%%%%%%%%%%%%%%%%%%%%%%%%%%%%%%%5

After searching through the HST archive we decided to use a set of
stellar population models instead of the one star from the archive
that matched our set-up (the same grism being the most important,
while size of the slit can be accounted for). Using single-metallicity
stellar population models of \citet{1999ApJ...513..224V} we
constructed a large stellar library from which to build the optimal
stellar template. Each galaxy long-slit spectrum was summed up along
the slit to make a higher $S/N$ spectrum, which was used to obtain the
optimal template. We also used additive Legendre polynomials to adapt
the continuum shape of the templates. This optimal template was then
used in the fit of the individual spectra along the slit.

A disadvantage of the Vazdekis models is that they are of lower
resolution than the STIS data. The FWHM of the Vazdekis library is 1.8
\AA\, compared to 0.8 \AA\, from STIS. This requires a degradation of
our data by $\sim$1.6 \AA. Although certain information is in this way
lost and the STIS spectral resolution is degraded, the smoothing of
the data helps in removing noise and the library ensures the extracted
kinematics do not suffer from an important systematic template
mismatch effect. Examples of constructed optimal templates are shown
in Fig.~\ref{f:spec}. The presented spectra are the sum of the central
10 rows, having a high $S/N$ which is needed for properly estimating
the optimal template. The overplotted dashed lines are our resulting
optimal templates for the galaxies, convolved with the determined
LOSVD of the galaxy. Typically a few (2-3) old-type stars from the
Vazdekis library were selected by the fitting routine for the optimal
template. The residuals between the galaxy and the optimal template
spectra are shown below each spectrum. The presented optimal templates
were used to extract the kinematics from the spatially binned
spectra. An alternative way would be to construct the optimal template
for each spatial bin and then use this to extract kinematics in the
same bin. This method is important for galaxies with stellar
populations changing between bins ($\sim0\farcs05$), but in this case
the low $S/N$ of the individual spectra do not justify this approach.

%%% Table 7.%%%%%%%%%%%%%%%%%%%%%%%%%%%%%%%%%%%%%%%%%%%%%%%%%%%%%%%%%%%%%%
\begin{table}

 \caption[]{Bins for kinematic extraction }

  \label{t:bin}
$$
  \begin{array}{ccrccc}
    \hline
   \hline
     \noalign{\smallskip}
     $galaxy$ & $bin$ & $r [arcsec]$ & $width [pix]$ & $range$ & $S/N$\\
    \noalign{\smallskip}
    \hline
   \hline
     \noalign{\smallskip}
     $NGC~4128$ & center  &  0.00  &  1 & 599-599 & 15\\
                & r1      &  0.05  &  1 & 600-600 & 14\\
                & r2      &  0.15  &  2 & 601-603 & 15\\
                & r3      &  0.25  &  3 & 604-607 & 13\\
                & r4      &  0.50  &  6 & 608-612 & 12\\
                & r5      &  1.25  & 24 & 613-637 & 12\\
                & l1      & -0.05  &  1 & 598-598 & 14\\
                & l2      & -0.10  &  1 & 596-597 & 12\\
                & l3      & -0.15  &  2 & 593-595 & 14\\
                & l4      & -0.30  &  3 & 589-592 & 13\\
                & l5      & -0.50  &  5 & 583-588 & 12\\
                & l6      & -0.95  & 13 & 569-582 & 12\\

    \hline
   \hline
     $NGC~4570$ & center  &  0.00  &  1 & 599-599 & 22\\
                & r1      &  0.05  &  1 & 600-600 & 21\\
                & r2      &  0.10  &  1 & 601-601 & 18\\
                & r3      &  0.15  &  2 & 602-604 & 20\\
                & r4      &  0.30  &  3 & 605-608 & 20\\
                & r5      &  0.50  &  5 & 609-614 & 19\\
                & r6      &  0.80  &  8 & 615-623 & 19\\
                & l1      & -0.05  &  1 & 598-598 & 19\\
                & l2      & -0.15  &  2 & 595-597 & 21\\
                & l3      & -0.25  &  3 & 591-594 & 21\\
                & l4      & -0.45  &  4 & 586-590 & 19\\
                & l5      & -0.70  &  6 & 579-585 & 18\\
  
     \hline
   \hline
     $NGC~4621$ & center  &  0.00  &  1 & 599-599 & 40\\
                & r1      &  0.05  &  1 & 600-600 & 34\\
                & r2      &  0.10  &  1 & 601-601 & 25\\
                & r3      &  0.20  &  2 & 602-604 & 26\\
                & r4      &  0.35  &  4 & 605-609 & 27\\
                & r5      &  0.65  &  8 & 610-618 & 26\\
                & l1      & -0.05  &  1 & 598-598 & 36\\
                & l2      & -0.10  &  1 & 597-597 & 28\\
                & l3      & -0.15  &  2 & 594-596 & 29\\
                & l4      & -0.30  &  3 & 590-593 & 26\\
                & l5      & -0.50  &  6 & 583-589 & 27\\
                & l6      & -0.95  & 11 & 571-582 & 25\\
        
    \hline
    \hline
     $NGC~5308$ & center  &  0.00  &  1 & 599-599 & 27\\
                & r1      &  0.05  &  1 & 600-600 & 23\\
                & r2      &  0.10  &  1 & 601-601 & 17\\
                & r3      &  0.20  &  2 & 602-604 & 17\\
                & r4      &  0.30  &  3 & 605-608 & 15\\
                & r5      &  0.55  &  6 & 609-615 & 15\\
                & l1      & -0.05  &  1 & 598-598 & 23\\
                & l2      & -0.10  &  1 & 597-597 & 17\\
                & l3      & -0.15  &  2 & 594-596 & 18\\
                & l4      & -0.30  &  3 & 590-593 & 17\\
                & l5      & -0.50  &  5 & 584-589 & 16\\
                & l6      & -0.90  & 11 & 572-583 & 15\\

    \noalign{\smallskip}
    \hline
   \hline
   \end{array}
$$

\end{table}
%%%%%%%%%%%%%%%%%%%%%%%%%%%%%%%%%%%%%%%%%%%%%%%%%%%%%%%%%%%%%%%%%%%%%%%%%

Table~\ref{t:bin} summarizes the details about the spatial bins used
for the extraction of kinematics. They were chosen after some
experimenting as a compromise between the $S/N$ and the spatial
resolution. The galaxies have different surface brightnesses and,
since the exposure times were similar, a unique scheme for all
galaxies was not useful. For each galaxy we assumed a target $S/N$ and
we binned accordingly. Generally, the spectra become too noisy to
measure the kinematics beyond 1\arcsec. In some cases, the central few
rows of spectra have $S/N$ high enough for extraction of the higher
order terms of LOSVD, but in general the $S/N$ is too low. Hence, we
decide to confine the extraction to only the first two moments
(assuming a Gaussian shape for LOSVD): mean stellar velocity ($V$) and
velocity dispersion ($\sigma$).

Another element which can heavily influence the results of the
extraction is specific to the spectral region of the
observations. \citet{2002AJ....124.2607B} compared the kinematics of a
number of galaxies extracted in two spectral regions: one around Mg$b$
lines and the other around Ca triplet. They found that if the
metallicities of the galaxies and template stars are not well matched
then direct template-fitting results are improved if the Mg$b$ lines
themselves are excluded from the fit and the velocity dispersion is
determined from the surrounding weaker lines. For galaxies with high
velocity dispersion this will be more important because of the
correlation between the velocity dispersion and the [Mg/Fe] ratio
\citep{1992ApJ...398...69W, 1998ApJS..116....1T, 2001MNRAS.323..615K},
which increases the strength of the Mg$b$ lines relative to the
surrounding Fe lines. Following the suggestion of
\citet{2002AJ....124.2607B} we also extracted kinematics excluding
from the fit the Mg$b$ lines (the excluded regions are shown on
Fig.~\ref{f:spec} as vertical lines crossing the spectra). When there
are significant differences between the two sets of extracted
kinematics we used the set obtained by excluding the Mg$b$ lines from
the fit for the further analysis and interpretations.

The errors were estimated using Monte-Carlo simulations. The LOSVD
parameters were derived from 100 realizations of the input spectrum,
where the value at each pixel is taken from a Gaussian distribution
with the mean of the initial spectrum and standard deviation given by
a robust sigma estimate of the residual of the fit to the initial
spectrum. Fig.~\ref{f:spec} shows an example of the residuals used to
estimate the standard deviation used in Monte-Carlo simulations (dots
under the spectra). All realizations provide a distribution of values
from which we estimate the 1$\sigma$ confidence limits. The values of
the extracted kinematics are presented in Tables~A.1-2 of Appendix A
and shown in Figs.~\ref{f:ngc4128} -- \ref{f:ngc5308}.

All galaxies except NGC~4621 show rather fast major axis
rotation. NGC~4621 is a special case with a previously discovered
counter-rotation in the center
(WEC02\nocite{2002A&A...396...73W}). There are some differences
between the kinematics extracted fitting the full spectral range and
excluding the Mg$b$ lines. They are the strongest for NGC~4621 and
NGC~4570. The somewhat larger error bars of the kinematic measurements
obtained not fitting the Mg$b$ region are the consequence of lowering
the $S/N$ by excluding the dominant spectral feature. We postpone
detailed description of all kinematic curves to
Section~\ref{s:discus}.

\subsection{Line strengths}

The spectral range of our observations is very limited covering only
the Mg$b$ and Fe5270 Lick/IDS indices (for definition of Lick/IDS
system and indices see \citet{1984ApJ...287..586B,
1994ApJS...94..687W, 1998ApJS..116....1T}). The red continuum pass
band of the Fe5270 index is truncated by the edge of our spectral
range and this index cannot be used in its defined form. A similar
case is found in Kuntschner et al. 2004 (in preparation) and
\citet{2004MNRAS.350...35F} where Fe5270 cannot be mapped over the
whole field-of-view of the integral-field spectrograph {\tt SAURON}
\citep{2001MNRAS.326...23B} due to the varying bandpass of the {\tt
SAURON} instrument. In their case, Kuntschner et al. (2004) redefine
the index to maximize the coverage of the field-of-view and retain the
sensitivity of the index towards changes in age, metallicity and
abundance ratios. The new index name is Fe5270s. It measures the same
spectral feature, but has a reduced spectral coverage in the red
pseudo-continuum band. The new index can be converted to the original
Lick/IDS system via the empirical formula \citep[Kuntschner et
al. 2004,][]{2004MNRAS.350...35F}:
\begin{displaymath}
\mbox{Fe}5270=1.26\times \mbox{Fe}5270s + 0.12
\end{displaymath}

\noindent The $1\sigma$ standard deviation of the above empirical
calibration is $\pm 0.05$ \AA\,for the Fe5270 index. More details on
the derivation of the new index and its relation to the standard
Lick/IDS index are given in Kuntschner et al. (2004).

%%%%% Table 8. %%%%%%%%%%%%%%%%%%%%%%%%%%%%%%%%%%%%%%%%%%%%%%%%%%%%%%%%%%%
\begin{table*}

 \caption[]{Line-strength indices measured in 0\farcs55 x 0\farcs2 aperture}
  \label{t:lsi}
$$
  \begin{array}{c|ccc|ccc|ccc|ccc}
    \hline
    \hline
    \noalign{\smallskip}

    \multicolumn{1}{c|}{$index$} &\multicolumn{3}{c|}{$NGC~4128$ } &
    \multicolumn{3}{c|}{$NGC~4570$}& \multicolumn{3}{c|}{$NGC~4621$}&
    \multicolumn{3}{c}{$NGC~5308$}\\ 

    $slit$& cen & neg & pos & cen & neg & pos & cen & neg & pos
    &cen & -neg & pos\\
    \noalign{\smallskip} \hline \hline \noalign{\smallskip}
  
   $ Mg$b    & 5.0 \pm 0.2 & -- & 5.0 \pm 0.4 & 4.5 \pm 0.2 & 2.9 \pm 0.3 & 3.7 \pm 0.4 & 5.9 \pm 0.1 & 6.6 \pm 0.3 & 4.7 \pm 0.3 & 4.8 \pm 0.1 & 5.5 \pm 0.5 & 5.8 \pm 0.6\\
   $ Fe5720$  & 2.9 \pm 0.1 & -- & 2.6 \pm 0.4 & 3.1 \pm 0.1 & 1.7 \pm 0.3 & 1.5 \pm 0.4 & 3.7 \pm 0.1 & 3.4 \pm 0.3 & 3.4 \pm 0.3 & 3.1 \pm 0.1 & 4.1 \pm 0.4 & 3.5 \pm 0.5\\

    \noalign{\smallskip} \hline \hline \noalign{\smallskip}

   $ Mg$b      & 4.8 \pm 0.2 & -- & 5.0 \pm 0.4 & 4.4 \pm 0.2 & 2.9 \pm 0.3 & 3.5 \pm 0.4 & 5.6 \pm 0.1 & 6.4 \pm 0.3 & 4.6 \pm 0.3 & 4.7 \pm 0.1 & 5.3 \pm 0.5 & 5.7 \pm 0.5\\
   $ Fe5720$  & 2.8 \pm 0.1 & -- & 2.6 \pm 0.4 & 3.1 \pm 0.1 & 1.7 \pm 0.3 & 1.4 \pm 0.4 & 3.6 \pm 0.1 & 3.4 \pm 0.3 & 3.3 \pm 0.3 & 3.0 \pm 0.1 & 4.0 \pm 0.5 & 3.5 \pm 0.4\\
   $B - V$  &0.97 & -- & 0.96  &--&--&--& 0.99 & 1.00 & 1.00 & 0.96 & 0.94 & 0.96\\

    \noalign{\smallskip}
    \hline
   \hline
   \end{array}
$$ 

{Notes -- First two rows in the table present line-strengths (in \AA)
corrected by the velocity dispersion measured using the whole spectral
region. The second two rows present line-strengths (in \AA) corrected
by the velocity dispersion measured excluding the Mg$b$ region from
the fit during extraction of kinematics. The last row presents B-V
color measured at the actual positions of slits within the same
slit-like aperture used for measuring line-strengths. The errors on
the color values are estimated to be 0.05 mag.}\looseness=-1

\end{table*}
%%%%%%%%%%%%%%%%%%%%%%%%%%%%%%%%%%%%%%%%%%%%%%%%%%%%%%%%%%%%%%%%%%%%%%%%%

Having this in mind we measured Fe5270s and Mg$b$ indices. The Mg$b$
index was measured using the Lick/IDS index definition, and all
spectra were first broadened to the resolution of the Lick/IDS
system. The Fe5270s index was later converted to index Fe5270 using
above relation. Unfortunately, we were not able to determine the
relevant offset to the Lick/IDS system, and correct for the
systematics, which come from differences in the continuum shape,
because there are no reference stars in the HST archive observed by
our and by the Lick/IDS instrumental setup. The size of the
corrections are probably similar to (or less than) our measurement
errors. To first order, as well as for determining the relative trends
in a galaxy, this is not very important, but has to be noted when
comparing with other studies.

Broadening of the lines by the velocity dispersion weakens most of the
lines and the index we measure must be corrected for this effect. This
can be achieved by determining an empirical correction factor
C($\sigma$)=index(0)/index($\sigma$) for a star observed with the same
instrumental setup. Index(0) is the index measured from the stellar
spectrum, $\sigma$ is the velocity dispersion of the LOSVD with which
the stellar spectrum is convolved and from which the index($\sigma$)
is measured. We used our unbroadened optimal template spectra to
calculate the index at $\sigma=0$ and at the corresponding velocity
dispersion, $\sigma$, of the galaxy spectrum. The proper correction
factor C($\sigma$) was then applied to both measured indices. We used
two approaches to extract kinematics and measure the velocity
dispersions (fitting the whole spectral region and excluding Mg$b$
region from the fit). If the measured velocity dispersions differ, the
velocity dispersion correction in the two cases will also be
different. We noted the difference applying both corrections on the
measured line-strengths.

We measured the Mg$b$ and Fe5270 indices from each spectral bin used
for kinematics. The corrected values and corresponding errors of the
index are presented in the Tables~A.1-2 and shown in the
Figs.~\ref{f:ngc4128} -- \ref{f:ngc5308}. The measured line-strengths
for galaxies with higher $S/N$ are relatively uniform with radius,
rising towards the center, with dips in the case of the nuclei of
NGC~4621 (Mg$b$) and NGC~5308 (Fe5720). NGC~4128 does not show any
trend, but rather a scatter of values, presumably due to the low
$S/N$, while in case of NGC~5308 Mg$b$ line-strengths are slightly
higher on one side of the galaxy. Generally, galaxies have high values
of Mg$b$ and Fe5270 indices. Detailed descriptions of spatially
resolved line-strengths for all galaxies are given in
Section~\ref{s:discus}.

\begin{figure} 
\centering
   \psfig{file=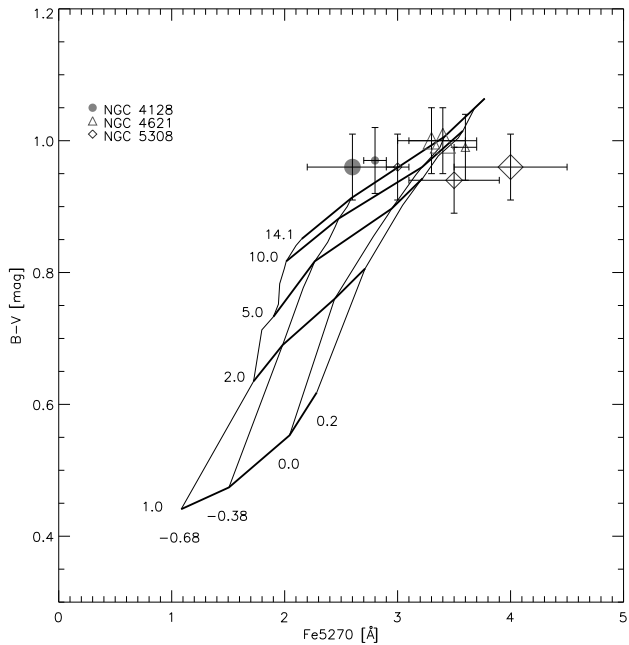, width=\columnwidth}
   \caption{\label{f:sspmg}Age/metallicity diagnostic diagram (B-V
   color vs. Fe5270 index). Horizontal thick solid lines are lines of
   constant age [Gyr] and vertical thin lines are lines of constant
   metallicity [Fe/H]) of \citet{1999ApJ...513..224V} models. The size
   of symbols is related to the position of the slit: the smallest
   symbols are for the {\it cen} slits, intermediate for the {\it pos}
   slits, and the biggest for the {\it neg} slits.}.  \centering
\end{figure}

The slits that were offset to the sides of the central slit do not
have the required $S/N$ to extract kinematics, but their summed
spectra can be used to determine the indices on the positions of the
galaxy outside the stellar disk. The binning of spectra is only useful
up to the point at which summing more spatial elements does not simply
add noise. We decided to use an aperture of $0\farcs55\times0\farcs2$
(summing up 5 rows on each side of the central row in the spectral
direction). With this approach the final side spectra used for the
measurements of line-strengths had at least $S/N$ $\approx 10$. In the
case of NGC~4128, however, from one side spectra we were not able to
extract any trustworthy measurement. Table~\ref{t:lsi} presents
line-strengths corrected for velocity dispersions measured by fitting
to the whole spectral range and excluding the Mg$b$ lines from the
fit. As it can be seen, the differences between the lines are
negligible and generally fall within the $1\sigma$ error bars. Only in
the case of the central slit of NGC~4621 which has the highest
velocity dispersion as well as $S/N$, there is a significant
difference. Adopting a conservative approach, we compared
line-strengths from the second two rows in Table~\ref{t:lsi} with the
stellar population synthesis models in Fig.~\ref{f:sspmg}.

We wish to compare the line-strengths measured on the disk with the
line-strengths measured on the bulge using the three slit
positions. The line-strength measurements on the summed spectra show
similarly high values as the spatially resolved measurements, although
with relatively lower values due to the smaller spatial
resolution. The Mg$b$ index values in NGC~4621 are particularly
high. Comparing with the literature we find similar values for
Mg$b$ and Fe 5270 index. Table~7 of \citet{1998ApJS..116....1T} list
values of the same indices for NGC~4621 and NGC~4570 (Mg$b$ 5.50 and
4.65\AA\,, Fe5270 3.59 and 3.49 \AA), which, keeping in mind the
unknown offset to the Lick system and the lower spatial resolution of
\citet{1998ApJS..116....1T} data (aperture of
$1\arcsec\times4\arcsec$), are in good agreement with our findings.

%%%%%%%%%%%Figure 7%%%%%%%%%%%%%%%%%%%%%%%%%%%%%%%%%%%%%%%%%%%%%%%%%%%%
%%%%%%%%%%%%%%%%%%%%%%%%%%%%%%%%%%%%%%%%%%%%%%%%%%%%%%%%%%%%%%%%%%%%%%%%%

Age and metallicity have similar effects on the integrated spectral
energy distributions that we measure from unresolved sources due to a
finely tuned conspiracy between age and metallicity variations
\citep{1994ApJS...95..107W}. Broad-band colors and many line-strength
indices are degenerate with respect to age and metallicity. This makes
the determination of the age and metallicities very difficult and
ideally one would like to use two indices which can break this
degeneracy. Usually, one or more Balmer lines ($H\beta$, $H\gamma$,
$H\delta$) are used as age indicators, and Mg$b$ or some Fe index
(Fe5270, Fe5335) as a metallicity indicator
\citep{1993PhDT.......172G, 1996ApJ...459..110F, 1998A&A...332...33M,
2000MNRAS.315..184K, 2000AJ....119.1645T}. The high index values of
our measured line-strengths also suggests the presence of non-solar
abundances of elements. If not properly treated, over-abundant indices
can give wrong age and metallicity estimates
\citep{2001MNRAS.323..615K}. A way around this issue is to define
metallicity indicators which are insensitive to abundance ratios
\citep{1993PhDT.......172G, 2003MNRAS.343..279T}. The preferred
indicator includes a combination of Mg$b$, Fe5335 and Fe5270 indices,
where Fe5270 is the least sensitive to changes of [$\alpha$/Fe]
abundance ratios \citep{2003MNRAS.343..279T}. We were not able to
construct such a metallicity indicator with the indices from our
spectral range, and we chose to use the least sensitive Fe5270 index
alone as a metallicity indicator. Since in our spectral range there
are no age indicators, we decided to use a combination of broad-band
B-V colors and Fe5270 index to construct an age/metallicity diagnostic
diagram (Fig.~\ref{f:sspmg}). The models presented by solid lines are
based on the \citet{1999ApJ...513..224V} single stellar population
models: color values were obtained from A.~Vazdekis web
site{\footnote{{\it http://www.iac.es/galeria/vazdekis/}}, while we
measured the Fe5270 index from the library spectra broadening them to
the Lick/IDS resolution. The combination of red colors and high
metallicities puts the measured points on Fig.~\ref{f:sspmg} on the
top right of the model grid, indicating old stellar populations and a
large spread of metallicities between the galaxies.

\section{Discussion}
\label{s:discus}
In the two previous sections we presented the observational results of
HST program 8667. They include photometric and spectroscopic
observations of four galaxies with nuclear stellar disks. Here we
analyze and discuss the observations.

\subsection{NGC~4128}
\label{ss:ngc4128}
The most distant galaxy in the sample is NGC~4128 (36 Mpc). It is an
S0 galaxy and it has not been detected in radio nor in IR. The
isophotal parameters show that it is disky between 35 and 530 pc. On 1
kpc scale it has boxy parameters and on larger scales it becomes disky
again.

The color image shows a red nucleus. The values for Fe5270 index
measured with an aperture bigger than the red nucleus are the smallest
in the sample of galaxies. This combination puts the points on
Fig.~\ref{f:sspmg} above the model grid. The difference in colors and
line-strengths between the two slit positions are small and within
errors indicate old stellar populations of $\sim$14 Gyr and
metallicities between [Fe/H]=-0.38 and solar.

It is probable that the difference in the color between the nucleus
and the rest of the galaxy, as well as the higher metallicity detected
in the nucleus is connected with the unusual spatially resolved
kinematic profiles (Fig.~\ref{f:ngc4128}). The velocity dispersion is
flat in the center. The velocity curve also shows an unusual
flattening in the central 0\farcs2, measurements being positive on the
both sides of the galaxy nucleus. Outside this radii the galaxy
rotates fast, as expected for a disk galaxy. Kinematics extracted
fitting to two different spectral regions are in a good agreement,
confirming the results. Having in mind the boxiness in the central
tens of arcseconds, the extracted kinematic indicates the existence of
a small ($\sim$35 pc in diameter) core, kinematically distinct from
the nuclear stellar disk. 

The B-V color profile on the last panel of Fig.~\ref{f:ngc4128} shows
a slightly shifted ($\sim$17 pc from the center) peak of the red
nucleus. This supports the presence of a distinct component in the
nucleus. On the other hand, the spatially resolved line-strengths do
not follow this trend. The spectral observations of this galaxy have
the smallest $S/N$ ratio, and the significance of this discovery is
just above 1$\sigma$. Deeper exposures of high spatial resolution,
preferably with an integral-field unit to cover the 2D structure, are
needed to confirm this result.

%%%%%%%%%%%Figure 7%%%%%%%%%%%%%%%%%%%%%%%%%%%%%%%%%%%%%%%%%%%%%%%%%%%%
\begin{figure}
\centering
   \psfig{file=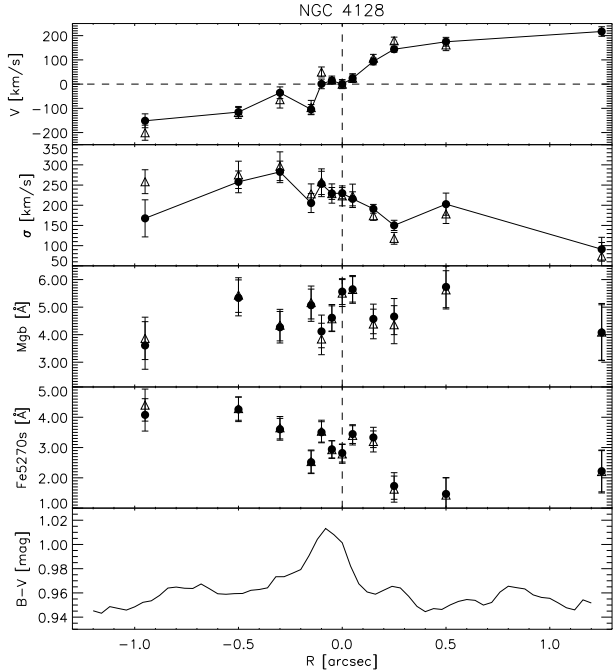, width=\columnwidth}
   \caption{\label{f:ngc4128}Kinematic and line-strength profiles for
   NGC~4128. From top to bottom: mean velocity, velocity dispersion,
   Mg$b$ index, Fe5270 index and B-V color profile. The color profile
   was extracted along the slit position averaging 0\farcs2
   perpendicular to the slit. The closed symbols represent measurement
   obtained by fitting the whole spectral region. The open symbols
   represent measurement by excluding the Mg$b$ line from the fit. }
\centering
\end{figure}

\subsection{NGC~4570}
\label{ss:ngc4570}

NGC~4570 is a well-studied galaxy with HST. The main result from
previous studies is that the inner region of the galaxy was shaped
under the influence of a weak bar \citep{1998MNRAS.298..267V}. The
colors reveal no difference between the disk and the bulge, and a
comparison with the stellar-population models indicate that the stars
in the galaxy are of intermediate age, but the FOS spectral data gave
a very high $H\beta$ line-strength suggesting recent star formation
(BJM98)\nocite{1998MNRAS.293..343V}. One of the questions raised by
these studies is whether all double-disk structures are the result of
bar-driven secular evolution.

To the previous photometric and spectroscopic observations we add new
spatially resolved spectroscopic measurements with STIS
(Fig.~\ref{f:ngc4570}). The probed region corresponds to the nuclear
disk and inner 2\arcsec. The velocity curve shows regular rotation
peaking at $\sim0\farcs15$ from the nucleus. The velocity dispersion
steeply rises and peaks in the center. The kinematic profiles are
similar to BJM98 ground based data, except the STIS data have a
steeper velocity curve and higher velocity dispersion. In contrast,
the FOS velocity dispersion from the same authors is about 50 km
s$^{-1}$ higher than STIS measurements; however, considering the error
bars of both measurement (their error on sigma is $\approx30$ km
s$^{-1}$) and the fact that our slit was significantly (for the width
of the slit) offset from the galaxy nucleus, these measurements can be
considered consistent with each other.

In the case of NGC~4570, the central velocity dispersion is somewhat
dependent on the spectral region used in the fit. Excluding Mg$b$
region systematically lowers the values by just over 1$\sigma$, but
increases the difference between this and BJM98 results.

%%%%%%%%%%%Figure 7%%%%%%%%%%%%%%%%%%%%%%%%%%%%%%%%%%%%%%%%%%%%%%%%%%%%
\begin{figure}
\centering
   \psfig{file=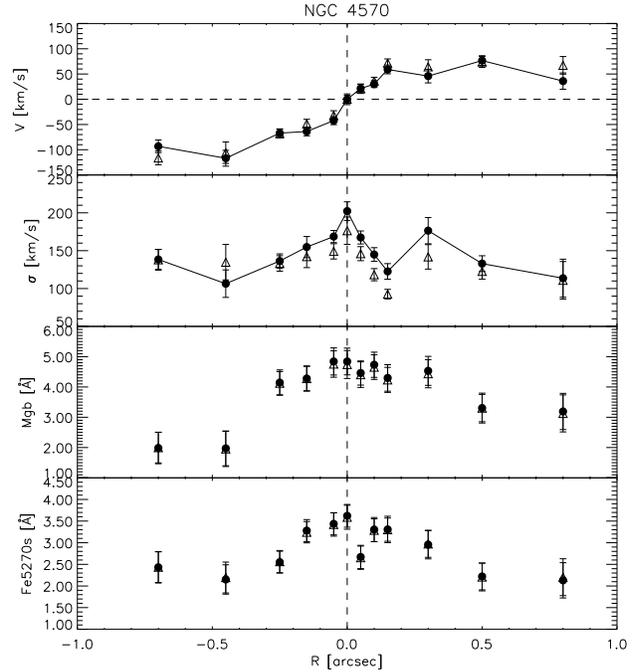, width=\columnwidth}
   \caption{\label{f:ngc4570}Kinematic and line-strength profiles for
   NGC~4570. From top to bottom: mean velocity, velocity dispersion,
   Mg$b$ index, Fe5270 index. The closed symbols represent measurement
   obtained by fitting the whole spectral region. The open symbols
   represent measurement by excluding the Mg$b$ line from the fit. }
\centering
\end{figure}

We also measure, within the errors, similar line-strengths to BMJ98,
but with higher spatial resolution and we can give an estimate of the
spatial changes in the indices. As can be seen from the
Fig.~\ref{f:ngc4570}, both measured indices show flattening in the
central 0\farcs5 ($\sim $60 pc). At larger radii the metallicity
drops. Also, the slits positioned on both side of the center measure
the smallest metallicity (Table~\ref{t:lsi}) in the sample and the
largest drop in values with respect to the center. This measurement
shows that the nuclear disk consists of different stellar populations
than the rest of the bulge, which is consistent with bar-driven
evolution.

\subsection{NGC~4621}
\label{ss:ngc4621}
The closest galaxy of the four is NGC~4621 (7 Mpc). It is also the
only galaxy classified as an elliptical and is the only galaxy from
the sample detected with IRAS (in the 12 $\mu$m band). In the
investigated range the galaxy is disky. The $b_{4}$ coefficient
steadily rises from the center to the distance of 140 pc when it
drops, but never reaching negative values. With increasing radius it
rises again, implying an outer disk.

The color images reveal the most interesting features in the
nucleus. The few central pixels are clearly much bluer than the rest
of the bulge (Table~\ref{t:color}). Another striking characteristic of
the galaxy, mentioned in section 2.3.2, is the extended blue and red
features visible on B-I (Fig.~\ref{f:nuc_disk}) and V-I images. The
blue feature makes an angle of 15\degr\, with the major axis. Although
the red feature spreads generally in the east-west direction (angle
with major axis is 103\degr), it is not as clearly defined as the blue
feature. We conclude, examining all color images, that the shape and
the extent of the two features can be interpreted as a blue disk-like
structure imbeded in the red bulge. The position angle of the blue
feature is unexpected for an axisymmetric galaxy with a nuclear
stellar disk. This significant structure perhaps can be explained
considering the kinematics of this galaxy.

%%%%%%%%%%%Figure 7%%%%%%%%%%%%%%%%%%%%%%%%%%%%%%%%%%%%%%%%%%%%%%%%%%%%
\begin{figure}
\centering
   \psfig{file=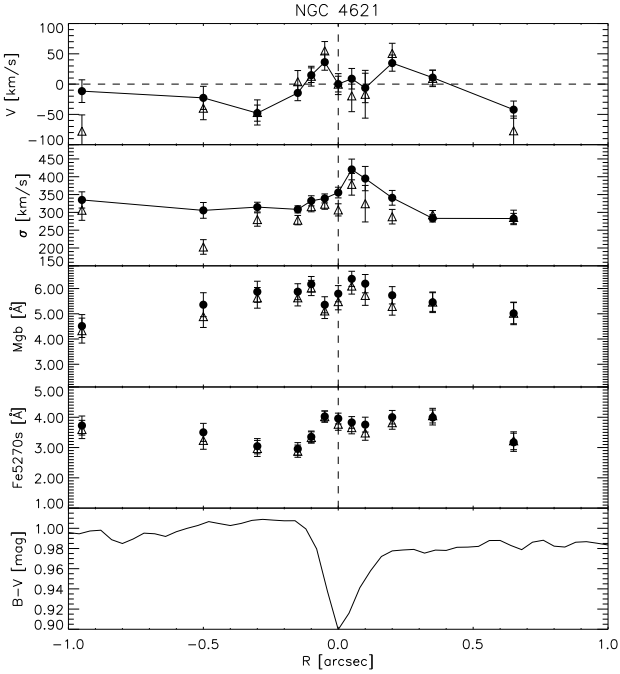, width=\columnwidth}
   \caption{\label{f:ngc4621}Kinematic and line-strength profiles for
   NGC~4621. From top to bottom: mean velocity, velocity dispersion,
   Mg$b$ index, Fe5270 index and B-V color profile. The color profile
   was extracted along the slit position averaging 0\farcs2
   perpendicular to the slit. The closed symbols represent measurement
   obtained by fitting the whole spectral region. The open symbols
   represent measurement by excluding the Mg$b$ line from the fit.}
\centering
\end{figure}

The velocity and velocity dispersion panels in Fig.~\ref{f:ngc4621}
clearly show the existence of a kinematically decoupled core
(KDC). This core was already detected by
WEC02\nocite{2002A&A...396...73W} who observed the galaxy with
integral field spectrograph OASIS mounted on CFHT and assisted by the
PUEO adaptive optics system. Complementing their OASIS observations,
WEC02\nocite{2002A&A...396...73W} also extracted kinematics from an
archival STIS observations in the Ca triplet region, showing a peak in
velocity dispersion 0\farcs05 from the center as well as confirming
the KDC with the same spatial resolution as in the data presented
here. The difference between the WEC02\nocite{2002A&A...396...73W} and
our kinematic profiles is largest in the velocity dispersion profile,
where our values lie systematically above the
WEC02\nocite{2002A&A...396...73W} measurements. This difference is
lower, but still present if we compare the kinematics extracted
excluding the Mg$b$ line from the fit ($\sim$60 kms$^{-1}$ for the
velocity dispersion peak, but with large error-bars in both
cases). This discrepancy could arise from the different slit positions
in the two studies, our slit being offset from the center and not
covering the KDC uniformly. The counter-rotation of the KDC could
lower the overall measured dispersion if the slit is placed over its
center, and, alternatively, if the slit misses the center of the KDC
the measured velocity dispersion will be higher.

The KDC on the WEC02\nocite{2002A&A...396...73W} OASIS data is not
aligned with the major axis and it has a similar position angle as the
blue feature on the color images presented here. Although the extent
of the KDC is smaller (total of $\sim2\arcsec$ or 60 pc) than the blue
feature on the B-I image, the existence of two structures could be
connected as a result of the same formation process.

Both Mg$b$ and Fe5270 line-strengths indices in NGC~4621 are the
highest in the sample, also suggesting the over-abundance ratios of
elements similar to trends for giant ellipticals
\citep{1998PhDT........24K, 2001MNRAS.323..615K}. Our metallicity
indicator, Fe5270, also has high values, with the central slit being
slightly more metal-rich than the side slits as well as
super-solar. Colors at the slit positions are red and the comparison
with the stellar population models indicates the age of the stars is
between 10 and 14 Gyr and the metallicity between solar and +0.2.

Our spatially resolved measurements of the indices, shown in
Fig.~\ref{f:ngc4621}, are higher than in previous studies
(e.g. \citet{2001MNRAS.323..615K} have central Mg$b$ $\sim5.21$ with
aperture of 3\farcs4), but our aperture is much smaller
($\sim0\farcs05$) than that of any previous study, and the values
outside the central arc-second approach the observed values from the
literature. The Mg$b$ index follows to some extent the changes in
colors, showing a small dip in the center, while this can not be said
for Fe5270 measurements.

The existence of the KDC and the blue features in the red bulge of the
NGC~4621 indicate two possible evolutionary scenarios. The visible
structures could be the result of a hierarchical formation scheme
\citep[e.g.][]{1994MNRAS.267..981K} involving a merger followed by a
starburst where the KDC is the remnant of the ejected stars that
later fell back in. These structures are relatively long lived, having
a relaxation time of $\sim$1 Gyr \citep{1987gady.book.....B}; however,
this is not long enough to explain the detected old age of the
stars. Alternatively, the structure could be produced by weak
bar-driven evolution, as in the case of NGC~4570
\citep{1998MNRAS.298..267V}, where the observed double disk structure
is the consequence of resonant frequencies in the galaxy, while the
blue feature and the KDC are the result of gas captured on retrograde
(`anomalous') orbits which are tilted with respect to the equatorial
plane \citep{1991A&A...252...75P, 1993IAUS..153..273F,
1997A&A...318L..39E}. Of course, a combination of both processes can
also lead to the present situation.

Distinguishing between the two scenarios is also difficult because the
galaxy has no obvious merger companion and it is nearly edge-on,
making the detection of a weak bar more difficult. There are other
cases of barred galaxies with similar properties to NGC~4621: i.e. an
edge-on system with double-disk structure, unusual photometric and
kinematic features and the difference in metallicity between the bulge
and the disk. An example of a similar, although boxier, galaxy with a
strong bar is NGC~7332. This galaxy was recently studied in detail by
\citet{2004MNRAS.350...35F}.  NGC~7332 is classified as an S0 galaxy
and has a double disk structure \citep{1996A&A...310...75S}. Examining
their {\tt SAURON} spectroscopic observations, Falc\'on-Barroso et
al. find a counter rotating stellar component within the central 250
pc. The galaxy also has complex gas morphology and the line-strength
maps show it is young everywhere. The authors conclude that NGC~7332
is an S0 galaxy with a bar viewed close to edge on. NGC~4621 and
NGC~7332 are similar in their morphologies, and, although different in
the stellar content, it is possible that NGC~4621 went through a
similar formation process as NGC~7332.

\subsection{NGC~5308}
\label{ss:ngc5308}
In many aspects NGC~5308 is different from the other galaxies in this
study. Our photometry reveals the largest nuclear stellar disk in the
sample of galaxies in this study. Unlike in the other galaxies, the
nuclear disk of NGC~5308 is very thin and bright. The diskiness
parameter, b$_{4}$, rises from the center and peaks at about 150 pc,
dropping to zero at $\sim$1 kpc and suggesting a distinction between
the two disks. At large radii the galaxy again becomes disky.

The stars in the disk of NGC~5308 rotate fast, reaching $\sim$100 km
s$^{-1}$ within 15 pc from the nucleus
(Fig.~\ref{f:ngc5308}). The velocity dispersion has a peak of about 300
km s$^{-1}$ in the center and is relatively flat in the inner 15
pc. This trend is also visible in the kinematics measured excluding
the Mg$b$ line from the fit, although the right hand side of the plot
shows considerably lower velocity dispersion values. This is reflected
in the panel with Mg$b$ values, which are slightly higher on the right
hand of the plot. The Mg$b$ and Fe5270 index values have opposite
trends. The small variations in the B-V color along the slit,
including the sudden blue dip in the center, are followed by the
line-strength measurements.

There is a big difference between the line-strengths measured on the
different slit positions (Table~\ref{t:lsi} and
Fig.~\ref{f:sspmg}). The nucleus, being just below solar metallicity,
is more metal-poor than the bulge which has a non-solar abundance
ratio of elements and the highest metallicity in the sample. The
nucleus and the investigated part of the bulge also have different
colors, with the center being redder. Comparing these results with the
stellar population synthesis models reveals an old stellar population
in the nucleus (14 Gyr), while the colors of the bulge suggest
intermediate age stellar component ($\sim$ 5-10 Gyr).

The color difference between the bulge and the nucleus is measured
because the side slits were positioned in the region of the minor axis
blue feature, especially visible on B-I image. This region is also
special for its high metal content. There are no hints of specific
morphological structures (such as a polar ring) along the minor axis
and the question is whether the rest of the bulge, especially the red
feature, and the nuclear stellar disk share the same
metallicity. Unfortunately our spectral observations did not cover the
necessary areas and we can only speculate on the processes that
created the observed structures.

If the metallicity of the minor axis blue feature and the nuclear disk
are the same, the formation of these two structures has to be
connected to the same formation scenario, most probably involving a
transportation of gas to the center, perhaps by a bar, for which
\citet{1996A&A...310...75S} found evidence in NGC~5308. If the
metallicity of the nuclear disk is lower than the metallicity of the
minor axis blue feature, but similar to the measured metallicity of
the nucleus, the two blue features in NGC~5308 were not created from
the same infalling material, but they still can be from the same
epoch. If the metallicity of the minor axis blue feature is equal to
the rest of the bulge, which is redder and therefore older, then the
younger stellar population in the blue feature must have been induced
by an internal process, perhaps ionization from the radiation
generated by the central black hole, which was turned on with the
infall of the material that made the blue disk, and later turned off
with the stabilization of the disk \citep{1994ApJ...432...52L}.

%%%%%%%%%%%Figure 7%%%%%%%%%%%%%%%%%%%%%%%%%%%%%%%%%%%%%%%%%%%%%%%%%%%%
\begin{figure}
\centering
   \psfig{file=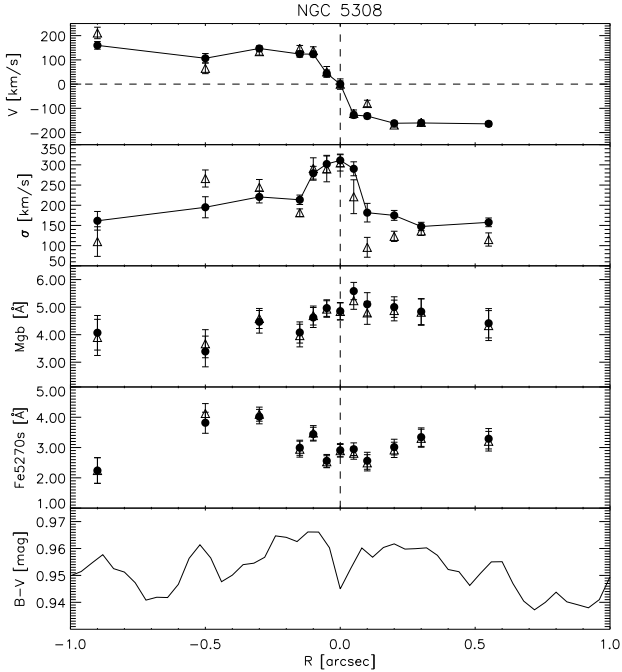,width=\columnwidth}
   \caption{\label{f:ngc5308}Kinematic and line-strength profiles for
   NGC~5308. From top to bottom: mean velocity, velocity dispersion,
   Mg$b$ index, Fe5270 index and B-V color profile. The color profile
   was extracted along the slit position averaging 0\farcs2
   perpendicular to the slit. The closed symbols represent measurement
   obtained by fitting the whole spectral region. The open symbols
   represent measurement by excluding the Mg$b$ line from the fit.}
\centering
\end{figure}

In section 2.3.2 we showed that the major axis blue feature in
NGC~5308 corresponds to the nuclear stellar disk. It is about 30 pc
thick which strongly suggest the galaxy is viewed very close to
edge-on. Comparing with the vertical scalelength of 34 edge-on spirals
presented by \citet{2002MNRAS.334..646K}, which are between 0.2 and
1.4 kpc thick, the disk in NGC~5308 is a remarkably thin disk. Note
that our estimate of the disk thickness can only be approximately
compared with the vertical scalelength measurements of Kregel et
al. It is also not possible to say much about the thickness of the
other nuclear disks, due to their inclination (not as edge-on as
NGC~5308). Whether this nuclear disk resembles the disks from the
group of ``super-thin'' galaxies, like UGC~7321 or IC~5249
\citep{1999AJ....118.2751M, 2000AJ....120.1764M, 2001A&A...379..374V},
is an open question. The sizes of the nuclear disk in NGC~5308 and
known ``super-thin'' disks are quite different as well as the
surrounding environment (stellar bulges and dark matter halos
respectively). A proper way to compare the disks is to measure the
radial and vertical sizes in a consistent manner, which is beyond the
scope of this paper.

The color and metallicity of the disk in NGC~5308 suggest the disk
could be made of a younger and more metal-poor stellar population than
the rest of the galaxy, implying it formed at a different epoch from
accreted material.

\section{Conclusions}
\label{s:con3}
We have presented photometric and spectroscopic observations of four
nearby early-type galaxies with nuclear stellar disks (NGC~4128,
NGC~4570, NGC~4621, NGC~5308). The observations consist of high
resolution images with WFPC2 using the F450W and F555W filters, and
STIS high resolution spectra through the 52x0.2 long-slit with the
G430M prism.

The photometric analysis reveals similarities and differences between
the galaxies. Nuclear stellar disks are clearly visible and are
photometrically disconnected from the large scale disks. NGC~4128
shows boxy isophotes on the inner and outer edge of the nuclear
disk. NGC~4621, the only E galaxy in the sample, is everywhere disky,
while NGC~5308 has a razor-thin ($\sim30$ pc) disk.

Color images reveal interesting and unexpected structures. NGC~4128
has a red nucleus, while NGC~4621 has a blue nucleus. Prominent color
features are visible on all galaxies. The blue feature in NGC~4128 is
analyzed in \ref{s:trans} and is the signature of a transient
event. NGC~4621 has a blue feature at an angle of $15\degr$ with the
major axis on top of a red bulge. It is likely connected to the KDC
discovered by WEC02\nocite{2002A&A...396...73W}. The nuclear stellar
disk in NGC~5308 is associated with the razor thin blue feature along
the major axis. NGC~5308 has another blue feature along the minor
axis. The colors of all three galaxies indicate old stellar
populations except for the bulge of NGC~5308 where the combination of
slightly less red colors and high metallicity lowers the age of the
stellar populations.

The high resolution spectroscopy was obtained at three positions on
each galaxy. One slit was positioned on the nuclear stellar disk, with
the PA equal to the major axis PA. Two additional slits were
positioned on both sides of the central slit, $\sim0\farcs3$ away from
the disk covering the bulge. The central slits were used to extract
spatially resolved kinematics and line-strengths. The $S/N$ permitted
the extraction of the mean stellar velocity and the velocity
dispersion, as well as the measurement of line-strengths.  The
kinematics will be used in a separate paper to estimate the black hole
masses in the centers of the galaxies.

Considering the shape of the spatially resolved kinematic curves, the
four galaxies could be sorted in two groups: fast and kinematically
disturbed rotators. NGC~4570 and NGC~5308 belong to the first
group. Their rotation curves show clear signature of the stellar
disks. The rotation curves of NGC~4128 and NGC~4621 are much more
complicated. In the case of NGC~4621 the unusual mean velocity and
velocity dispersion curves are consistent with the known KDC
(WEC02\nocite{2002A&A...396...73W}) in the nucleus. Although based on
a $1\sigma$ detection, we report the discovery of a similar
kinematically distinct core in the case of NGC~4128.

Spatially resolved line-strength measurements along the disk indicate
that all four galaxies are more metal-rich in the inner 0\farcs5 than
outside this radius. Both measured indices (Mg$b$ and Fe5270) increase
towards the center, except in the case of NGC~5308 where Fe5270 has an
opposite trend to Mg$b$ index. Non-solar abundance ratios of [Mg/Fe],
hinted by results of the extraction of kinematics, are present in
NGC~4570 and NGC~4621, and to some extent also in NGC~5308.

The slits positioned on the bulges had low $S/N$ and no spatially
resolved kinematics were extracted. However, by binning the spectra,
it was possible to measure the line-strengths at one position and
compare them to the values of the nuclei. The objects show various
structure: NGC~4128 has similar metallicities at the different slit
positions, NGC~4570 and NGC~4621 have higher metallicity in the
nucleus, while NGC~5308 in the bulge. Generally, the galaxies show a
spread in metallicity from sub- to super-solar.

This study shows the diversity within this class of objects, but also
emphasizes the similarities in the photometry and kinematics. The red
color gradient in the nuclei of NGC~4128 and the blue features in
NGC~4621 and NGC~5308 suggest the existence of different stellar
populations on small scales ($\sim$100-500 pc). The investigated
galaxies were chosen as galaxies with specific nuclear morphologies:
nuclear stellar disks. However, except in the case of NGC~5308 the
colors of the disks are not much different from the bulge, as
previously noted by \citet{1997ApJ...481..710C}. The existence of
other color features is a surprise. In two galaxies (NGC~4128 and
NGC~4621) these color features are followed by the existence of a
KDC. The other two galaxies do not show any peculiarities in their
kinematics. Also, if the KDC in NGC~4621 is connected to the
misaligned blue feature, we can conclude, similar to
\citet{1997ApJ...491..545C}, that KDCs are not kinematic counterparts
of the nuclear stellar disks. This gives credit to the complexity of
formation scenarios that demands a separate study per galaxy, but
there are a few most likely frameworks, outlined also in
\citet{1998MNRAS.300..469S}, in which the processes responsible for
the observed structures operate.

The formation of nuclear disks, rings and double disk structures in
early-type galaxies can be explained through secular evolution driven
by weak bars as shown by \citet{1996A&A...312..777E} and
\citet{1998MNRAS.298..267V} in the cases of M~104 and NGC~4570,
respectively. This mechanism, through the evolution of the bar,
explains the double-disk morphology. Support for this scenario comes
from the fact that S0 galaxies have high line-strengths
\citep{1996ApJ...459..110F} and there are evidences of embedded bars
in early-type galaxies \citep[e.g. M104, NGC~4570,
NGC~7332,][]{1998A&AS..131..265S}. This model of bar-driven evolution
is consistent with the observations in the presented galaxies, even in
the cases of the galaxies with KDCs, such as NGC~4621, but also
NGC~4128, which is additionally boxy and presents an interesting
case. The time varying triaxial potential of the bars offers exotic
orbits that could explain the existence of kinematic and photometric
features. In this scenario, the KDCs are created from enriched
material transported inwards (perhaps even gas acquired through a
merger), which gets frozen on retrograde orbits tilted with the
respect to the equatorial plane.

Other possibilities involve a merger scenario (capture of gas that
settles in the principle plane forming stars, and/or makes tidal
inflows that create KDCs), or growth of a central black hole
\citep{1994ApJ...432...52L}. A black hole stabilizes the disk and
within this scenario a connection to quasars can be made by stopping
the fueling of the central engines with the formation of a stable
disk. None of the previously investigated nuclear stellar disk
galaxies has an active nucleus, although they do harbor
$10^{8-9}_{\Msun}$ black holes \citep{1996ApJ...459L..57K,
1996ApJ...473L..91K, 1999ApJ...514..704C}. This makes them descendants
of quasars that spent their fuel (there is not much dust or gas in
most of these galaxies), or quasars that, through dynamical evolution,
turned off the central engine (stabilization of the disk due to the
growth of the black hole, disappearance of bars that transport the
material to the center).

Nuclear disks are easier to find in edge-on systems; however, the
influence of weak bars is correspondingly more difficult to
ascertain. Detailed spectroscopic studies with two dimensional
coverage of the major features (nuclei, stellar disks, KDC,
photometric features) are necessary to chose between the present
formation scenarios. A careful investigation of the two-dimensional
kinematic properties and their connection to the distribution of
line-strengths (metal content and age of stellar populations) can
offer decisive tools to deduce the nature and nurture of galaxies with
nuclear stellar disks.

\begin{acknowledgements}
We are grateful to Michele Cappellari, Eric Emsellem, Richard
McDermid, Gijs Verdoes Kleijn, Frank van den Bosch, Zlatan Tsvetanov and Tim
de Zeeuw for comments and discussions. DK thanks Michele Cappellari
and Harald Kuntschner for making available the pPXF and
line-strengths measurement software, respectively. This research has
made use of the NASA/IPAC Extragalactic Database (NED) which is
operated by the Jet Propulsion Laboratory, California Institute of
Technology, under contract with the National Aeronautics and Space
Administration. This work also used LEDA database. DK was supported by
NOVA, the Netherlands Research school for Astronomy.
\end{acknowledgements}

%----------------------------------------------------------------

%%%%%%%%%%%%%%%%%%%%%%%%%%%%%%%%%%%%%%%%%%%%%%%%%%%%%%%%%%%%%%%%%%%%%%%%%%%
% Reference List                                                          %
%%%%%%%%%%%%%%%%%%%%%%%%%%%%%%%%%%%%%%%%%%%%%%%%%%%%%%%%%%%%%%%%%%%%%%%%%%%
\bibliographystyle{aa}
%\bibliography{../refs.bib}
\bibliography{nuclear_disks_astroph.bbl}

%%%%%%%%%%%%%%%
% Appendices
%%%%%%%%%%%%%%%

\clearpage
\appendix

\section{Extracted kinematics}
\label{s:extrkin}

%%% Table A.1%%%%%%%%%%%%%%%%%%%%%%%%%%%%%%%%%%%%%%%%%%%%%%%%%%%%%%%%%%%%%
\begin{table}[!h]

\caption{ Measured kinematics and line-strengths for observed
 galaxies using full spectral region in the fit}
  \label{t:A1}
\vspace{-0.8cm}
$$
  \begin{array}{rcccccccc}
    \hline
    \hline
    \noalign{\smallskip}
    $radius$ & $V$ & \delta V & \sigma &\delta \sigma& $Mg$b &
    \delta$Mg$b & $Fe5270$ & \delta$Fe5270$\\
    (1)&(2)&(3)&(4)&(5)&(6)&(7)&(8)&(9)\\
   \noalign{\smallskip}
    \hline
    \hline 
    $NGC~4128$\\
    \hline 
    \hline 
 
              -0.95  &  2323. & 29. & 168. & 46. & 3.6 & 0.9 & 4.1 & 0.5\\[-0.6mm]
              -0.50  &  2359. & 19. & 258. & 27. & 5.3 & 0.7 & 4.3 & 0.4\\[-0.6mm]
              -0.30  &  2438. & 24. & 283. & 27. & 4.3 & 0.6 & 3.6 & 0.4\\[-0.6mm]
              -0.15  &  2370. & 20. & 206. & 24. & 5.1 & 0.6 & 2.5 & 0.4\\[-0.6mm]
              -0.10  &  2475. & 20. & 253. & 31. & 4.1 & 0.6 & 3.5 & 0.3\\[-0.6mm]
              -0.05  &  2489. & 12. & 229. & 15. & 4.6 & 0.5 & 2.9 & 0.3\\[-0.6mm]
               0.00  &  2474. & 13. & 230. & 14. & 5.6 & 0.5 & 2.8 & 0.3\\[-0.6mm]
               0.05  &  2498. & 14. & 216. & 18. & 5.6 & 0.5 & 3.4 & 0.3\\[-0.6mm]
               0.15  &  2568. & 14. & 191. & 11. & 4.6 & 0.5 & 3.3 & 0.3\\[-0.6mm]
               0.25  &  2618. & 12. & 150. & 13. & 4.7 & 0.7 & 1.7 & 0.4\\[-0.6mm]
               0.50  &  2649. & 18. & 203. & 27. & 5.7 & 0.8 & 1.5 & 0.5\\[-0.6mm]
               1.25  &  2691. & 20. &  91. & 30. & 4.1 & 1.0 & 2.2 & 0.7\\[-0.6mm]

    \hline  
    \hline 
   $NGC~4570$\\
    \hline 
    \hline 
              -0.70 & 1827. & 13. & 138. & 13. & 2.0 & 0.5 & 2.4 & 0.4\\[-0.6mm]
              -0.45 & 1803. & 16. & 106. & 18. & 2.0 & 0.6 & 2.2 & 0.3\\[-0.6mm]
              -0.25 & 1853. &  9. & 136. &  9. & 4.1 & 0.4 & 2.6 & 0.3\\[-0.6mm]
              -0.15 & 1856. &  9. & 155. & 14. & 4.3 & 0.4 & 3.3 & 0.3\\[-0.6mm]
              -0.05 & 1877. &  8. & 169. &  8. & 4.8 & 0.4 & 3.4 & 0.3\\[-0.6mm]
               0.00 & 1920. &  8. & 202. & 12. & 4.8 & 0.4 & 3.6 & 0.3\\[-0.6mm]
               0.05 & 1941. &  8. & 167. &  9. & 4.5 & 0.4 & 2.7 & 0.3\\[-0.6mm]
               0.10 & 1950. &  7. & 145. &  9. & 4.7 & 0.4 & 3.3 & 0.3\\[-0.6mm]
               0.15 & 1979. &  7. & 123. & 10. & 4.3 & 0.4 & 3.3 & 0.3\\[-0.6mm]
               0.30 & 1966. & 13. & 176. & 17. & 4.5 & 0.5 & 3.0 & 0.3\\[-0.6mm]
               0.50 & 1996. & 10. & 133. & 10. & 3.3 & 0.5 & 2.2 & 0.3\\[-0.6mm]
               0.80 & 1956. & 16. & 114. & 25. & 3.2 & 0.6 & 2.1 & 0.4\\[-0.6mm]

    \hline
    \hline 
    $NGC~4621$\\
    \hline 
    \hline 
              -0.95 & 583. & 19. & 335. & 23. & 4.5 & 0.5 & 3.7 & 0.3\\[-0.6mm]
              -0.50 & 572. & 19. & 306. & 22. & 5.4 & 0.5 & 3.5 & 0.3\\[-0.6mm]
              -0.30 & 546. & 13. & 315. & 14. & 5.9 & 0.4 & 3.0 & 0.3\\[-0.6mm]
              -0.15 & 580. & 13. & 308. & 10. & 5.9 & 0.3 & 3.0 & 0.2\\[-0.6mm]
              -0.10 & 609. & 12. & 332. & 14. & 6.2 & 0.3 & 3.4 & 0.2\\[-0.6mm]
              -0.05 & 631. & 13. & 339. & 13. & 5.4 & 0.3 & 4.0 & 0.2\\[-0.6mm]
               0.00 & 594. & 13. & 356. & 15. & 5.8 & 0.3 & 4.0 & 0.2\\[-0.6mm]
               0.05 & 603. & 17. & 420. & 29. & 6.4 & 0.3 & 3.8 & 0.2\\[-0.6mm]
               0.10 & 588. & 24. & 395. & 34. & 6.2 & 0.4 & 3.8 & 0.2\\[-0.6mm]
               0.20 & 629. & 13. & 341. & 21. & 5.7 & 0.3 & 4.0 & 0.2\\[-0.6mm]
               0.35 & 605. & 12. & 283. & 10. & 5.5 & 0.4 & 4.0 & 0.2\\[-0.6mm]
               0.65 & 552. & 14. & 283. & 14. & 5.0 & 0.4 & 3.2 & 0.3\\[-0.6mm]

    \hline
    \hline 
    $NGC~5308$\\
    \hline 
    \hline 
              -0.90 & 2295. & 16. & 162. & 23. & 4.1 & 0.6 & 2.2 & 0.4\\[-0.6mm]
              -0.50 & 2241. & 20. & 195. & 26. & 3.4 & 0.6 & 3.8 & 0.3\\[-0.6mm]
              -0.30 & 2282. & 11. & 221. & 15. & 4.5 & 0.4 & 4.0 & 0.2\\[-0.6mm]
              -0.15 & 2259. & 13. & 214. & 12. & 4.1 & 0.4 & 3.0 & 0.3\\[-0.6mm]
              -0.10 & 2258. & 12. & 279. & 17. & 4.6 & 0.4 & 3.4 & 0.2\\[-0.6mm]
              -0.05 & 2179. & 14. & 302. & 21. & 5.0 & 0.3 & 2.6 & 0.2\\[-0.6mm]
               0.00 & 2135. & 13. & 311. & 16. & 4.9 & 0.3 & 2.9 & 0.2\\[-0.6mm]
               0.05 & 2007. & 12. & 290. & 17. & 5.6 & 0.3 & 2.9 & 0.2\\[-0.6mm]
               0.10 & 2003. & 11. & 182. & 23. & 5.1 & 0.4 & 2.6 & 0.3\\[-0.6mm]
               0.20 & 1973. &  9. & 175. & 12. & 5.0 & 0.4 & 3.0 & 0.3\\[-0.6mm]
               0.30 & 1974. &  8. & 147. & 10. & 4.8 & 0.5 & 3.3 & 0.3\\[-0.6mm]
               0.55 & 1971. & 10. & 158. & 11. & 4.4 & 0.5 & 3.3 & 0.3\\[-0.6mm]
  \noalign{\smallskip}
    \hline
    \hline
  \end{array}
$$ 

%\vspace{-0.3cm} 
{Notes -- Col.~(1): radius of
measurement (arcsec); Col.~(2): mean velocity ($\kms$); Col.~(3) mean
velocity error ($\kms$); Col.~(4): velocity dispersion ($\kms$);
Col.~(5): velocity dispersion error ($\kms$); Col.~(6): Mg$b$ index
(\AA); Col.~(7) Mg$b$ index error (\AA); Col.~(8) Fe5270 index
(\AA);Col.~(9) Fe5270 index error (\AA)\looseness=-1}

\end{table}
%%%%%%%%%%%%%%%%%%%%%%%%%%%%%%%%%%%%%%%%%%%%%%%%%%%%%%%%%%%%%%%%%%%%%%%%%%%%

%%% Table A.2.%%%%%%%%%%%%%%%%%%%%%%%%%%%%%%%%%%%%%%%%%%%%%%%%%%%%%%%%%%%%%%
\begin{table}[!h]

 \caption[]{Measured kinematics and line-strengths for observed
 galaxies excluding Mg$b$ region from the fit}
  \label{t:A2}
%\vspace{-0.7cm}
$$
  \begin{array}{rcccccccc}
    \hline
    \hline
    \noalign{\smallskip}
    $radius$ & $V$ & \delta V & \sigma &\delta \sigma& $Mg$b &
    \delta$Mg$b & $Fe5270$ & \delta$Fe5270$\\
    (1)&(2)&(3)&(4)&(5)&(6)&(7)&(8)&(9)\\
   \noalign{\smallskip}
    \hline
    \hline 
    $NGC~4128$\\
    \hline 
    \hline 
              -0.95 & 2294. & 31. & 258. & 30. & 3.9 & 0.8 & 4.4 & 0.5\\[-0.5mm]
              -0.50 & 2378. & 25. & 275. & 34. & 5.4 & 0.6 & 4.3 & 0.4\\[-0.5mm]
              -0.30 & 2431. & 35. & 297. & 36. & 4.3 & 0.6 & 3.7 & 0.4\\[-0.5mm]
              -0.15 & 2398. & 30. & 228. & 25. & 5.1 & 0.6 & 2.5 & 0.4\\[-0.5mm]
              -0.10 & 2545. & 21. & 258. & 32. & 3.8 & 0.6 & 3.5 & 0.4\\[-0.5mm]
              -0.05 & 2512. & 19. & 229. & 24. & 4.6 & 0.5 & 2.9 & 0.3\\[-0.5mm]
               0.00 & 2495. & 19. & 223. & 25. & 5.5 & 0.5 & 2.8 & 0.3\\[-0.5mm]
               0.05 & 2520. & 19. & 223. & 29. & 5.6 & 0.5 & 3.4 & 0.3\\[-0.5mm]
               0.15 & 2603. & 15. & 174. & 12. & 4.4 & 0.5 & 3.2 & 0.3\\[-0.5mm]
               0.25 & 2674. & 15. & 118. & 15. & 4.4 & 0.7 & 1.6 & 0.4\\[-0.5mm]
               0.50 & 2657. & 24. & 179. & 24. & 5.6 & 0.7 & 1.4 & 0.6\\[-0.5mm]
               1.25 & 2688. & 23. &  74. & 33. & 4.0 & 1.0 & 2.2 & 0.7\\[-0.5mm]
    \hline
    \hline 
    $NGC~4570$\\
    \hline 
    \hline 
              -0.70 & 1805. & 14. & 138. & 14. & 2.0 & 0.5 & 2.4 & 0.4\\[-0.5mm]    
              -0.45 & 1815. & 21. & 131. & 23. & 2.0 & 0.6 & 2.2 & 0.4\\[-0.5mm]    
              -0.25 & 1853. &  8. & 133. & 10. & 4.1 & 0.4 & 2.6 & 0.3\\[-0.5mm]    
              -0.15 & 1872. &  9. & 142. & 15. & 4.2 & 0.4 & 3.2 & 0.2\\[-0.5mm]    
              -0.05 & 1889. &  9. & 149. & 10. & 4.8 & 0.4 & 3.4 & 0.3\\[-0.5mm]    
               0.00 & 1921. & 11. & 176. & 18. & 4.7 & 0.5 & 3.6 & 0.3\\[-0.5mm]    
               0.05 & 1942. &  9. & 146. &  9. & 4.4 & 0.4 & 2.7 & 0.3\\[-0.5mm]    
               0.10 & 1957. &  8. & 118. &  8. & 4.7 & 0.4 & 3.3 & 0.3\\[-0.5mm]    
               0.15 & 1993. &  8. &  93. &  6. & 4.2 & 0.4 & 3.3 & 0.3\\[-0.5mm]    
               0.30 & 1985. & 14. & 142. & 16. & 4.4 & 0.5 & 3.0 & 0.3\\[-0.5mm]    
               0.50 & 1995. & 11. & 123. & 10. & 3.3 & 0.5 & 2.2 & 0.3\\[-0.5mm]
               0.80 & 1988. & 17. & 111. & 25. & 3.1 & 0.6 & 2.2 & 0.4\\[-0.5mm]
 
    \hline
    \hline 
    $NGC~4621$\\
    \hline 
    \hline 
              -0.95 & 531. & 27. & 306. & 29. & 4.3 & 0.5 & 3.6 & 0.3\\[-0.5mm]
              -0.50 & 569. & 19. & 203. & 20. & 4.9 & 0.4 & 3.2 & 0.3\\[-0.5mm]
              -0.30 & 562. & 21. & 280. & 19. & 5.6 & 0.4 & 3.0 & 0.3\\[-0.5mm]
              -0.15 & 612. & 18. & 278. & 13. & 5.6 & 0.3 & 2.9 & 0.2\\[-0.5mm]
              -0.10 & 622. & 16. & 316. & 15. & 6.0 & 0.3 & 3.3 & 0.2\\[-0.5mm]
              -0.05 & 663. & 15. & 323. & 14. & 5.1 & 0.3 & 4.0 & 0.2\\[-0.5mm]
               0.00 & 608. & 17. & 307. & 17. & 5.5 & 0.3 & 3.8 & 0.2\\[-0.5mm]
               0.05 & 589. & 26. & 379. & 30. & 6.1 & 0.3 & 3.7 & 0.2\\[-0.5mm]
               0.10 & 592. & 39. & 324. & 51. & 5.7 & 0.4 & 3.5 & 0.2\\[-0.5mm]
               0.20 & 659. & 17. & 288. & 20. & 5.3 & 0.3 & 3.8 & 0.2\\[-0.5mm]
               0.35 & 618. & 14. & 293. & 12. & 5.5 & 0.4 & 4.1 & 0.2\\[-0.5mm]
               0.65 & 531. & 24. & 286. & 20. & 5.0 & 0.4 & 3.2 & 0.3\\[-0.5mm]
    \hline
    \hline 
    $NGC~5308$\\
    \hline 
    \hline 
              -0.90 & 2333. & 24. & 110. & 37. & 3.9 & 0.7 & 2.2 & 0.4\\[-0.5mm]
              -0.50 & 2188. & 22. & 266. & 21. & 3.7 & 0.5 & 4.1 & 0.3\\[-0.5mm]
              -0.30 & 2256. & 14. & 244. & 19. & 4.6 & 0.4 & 4.1 & 0.2\\[-0.5mm]
              -0.15 & 2270. & 12. & 181. & 10. & 4.0 & 0.4 & 2.9 & 0.3\\[-0.5mm]
              -0.10 & 2261. & 16. & 291. & 26. & 4.7 & 0.3 & 3.5 & 0.2\\[-0.5mm]
              -0.05 & 2173. & 22. & 290. & 32. & 4.9 & 0.3 & 2.5 & 0.2\\[-0.5mm]
               0.00 & 2122. & 21. & 305. & 20. & 4.8 & 0.3 & 2.9 & 0.2\\[-0.5mm]
               0.05 & 2001. & 14. & 221. & 42. & 5.2 & 0.3 & 2.8 & 0.2\\[-0.5mm]
               0.10 & 2042. & 13. &  96. & 25. & 4.8 & 0.4 & 2.5 & 0.3\\[-0.5mm]
               0.20 & 1954. &  9. & 123. & 13. & 4.9 & 0.4 & 2.9 & 0.3\\[-0.5mm]
               0.30 & 1965. &  9. & 136. & 11. & 4.8 & 0.5 & 3.3 & 0.3\\[-0.5mm]
               0.55 & 1955. & 10. & 115. & 16. & 4.3 & 0.5 & 3.2 & 0.3\\[-0.5mm]
 
  \noalign{\smallskip}
    \hline
    \hline
  \end{array}
$$ 
%{Notes -- Col.~(1): radius of measurement (arcsec); Col.~(2): mean
%velocity (km/s); Col.~(3) mean velocity error (km/s); Col.~(4):
%velocity dispersion (km/s); Col.~(5): velocity dispersion error
%(km/s); Col.~(6): Mgb index (\AA); Col.~(7) Mgb index error (\AA);
%Col.~(8) Fe5270s index (\AA);Col.~(9) Fe5270s index error (\AA);}
{Notes -- same as in Table~\ref{t:A1}.}
\end{table}
%%%%%%%%%%%%%%%%%%%%%%%%%%%%%%%%%%%%%%%%%%%%%%%%%%%%%%%%%%%%%%%%%%%%%%%%%%
 
\clearpage

\section{A transient in NGC~4128}
\label{s:trans}
Examining the HST -- WFPC2 images of the nearby S0 galaxy NGC~4128,
taken on May 14 2001 in the F450W and F555W filters, a peculiar source
of light was discovered very close to the center of the galaxy
(Fig.~\ref{f:super}). The source is offset by 0\farcs14 west and
0\farcs32 north of the galaxy center on the WFPC2 images. In the HST
archive there are images of the galaxy taken on March 17 1999, with
filter F702W, observed under PID 6357 (PI Jaffe), which do not show
any additional light source next to the nucleus of the galaxy. This
clearly indicates the appearance of a transient object in our data. In
this Appendix we try to deduce the origin of the
transient. Section~\ref{ss:extr} presents a new method of extraction
of the light contribution from the background galaxy and the
photometric analysis, followed by discussion on the nature of the
object in section~\ref{ss:dico}. \looseness=-1

\subsection{Extraction of light and results}
\label{ss:extr}
The general properties of NGC~4128 were described in the main
text. Here we add that the dust in the galaxy has not been detected
\citep{2001AJ....121.2928T}. This ensures there is no significant
extinction although the object is almost in the center of the
galaxy. The galactic component of the reddening in the direction of
the NGC\,4128 is estimated to be $E(B-V)=$ 0.02 mag
\citep{1998ApJ...500..525S}.

%%%%%%%%%%%Figure 7%%%%%%%%%%%%%%%%%%%%%%%%%%%%%%%%%%%%%%%%%%%%%%%%%%%%
\begin{figure} %\centering
{\hbox { \epsfxsize=0.49\columnwidth%\textwidth
\epsfbox{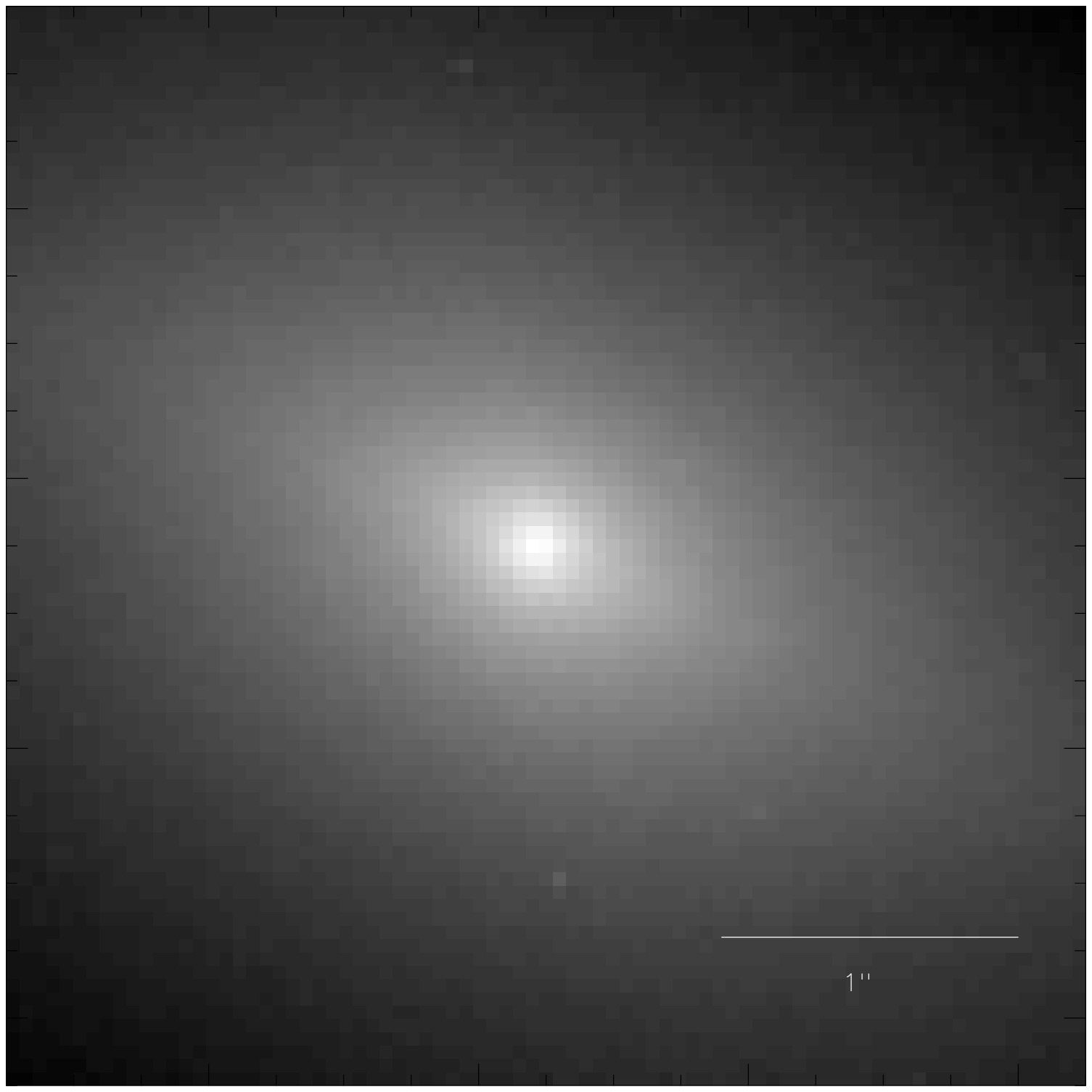}
\epsfxsize=0.49\columnwidth%\textwidth
\epsfbox{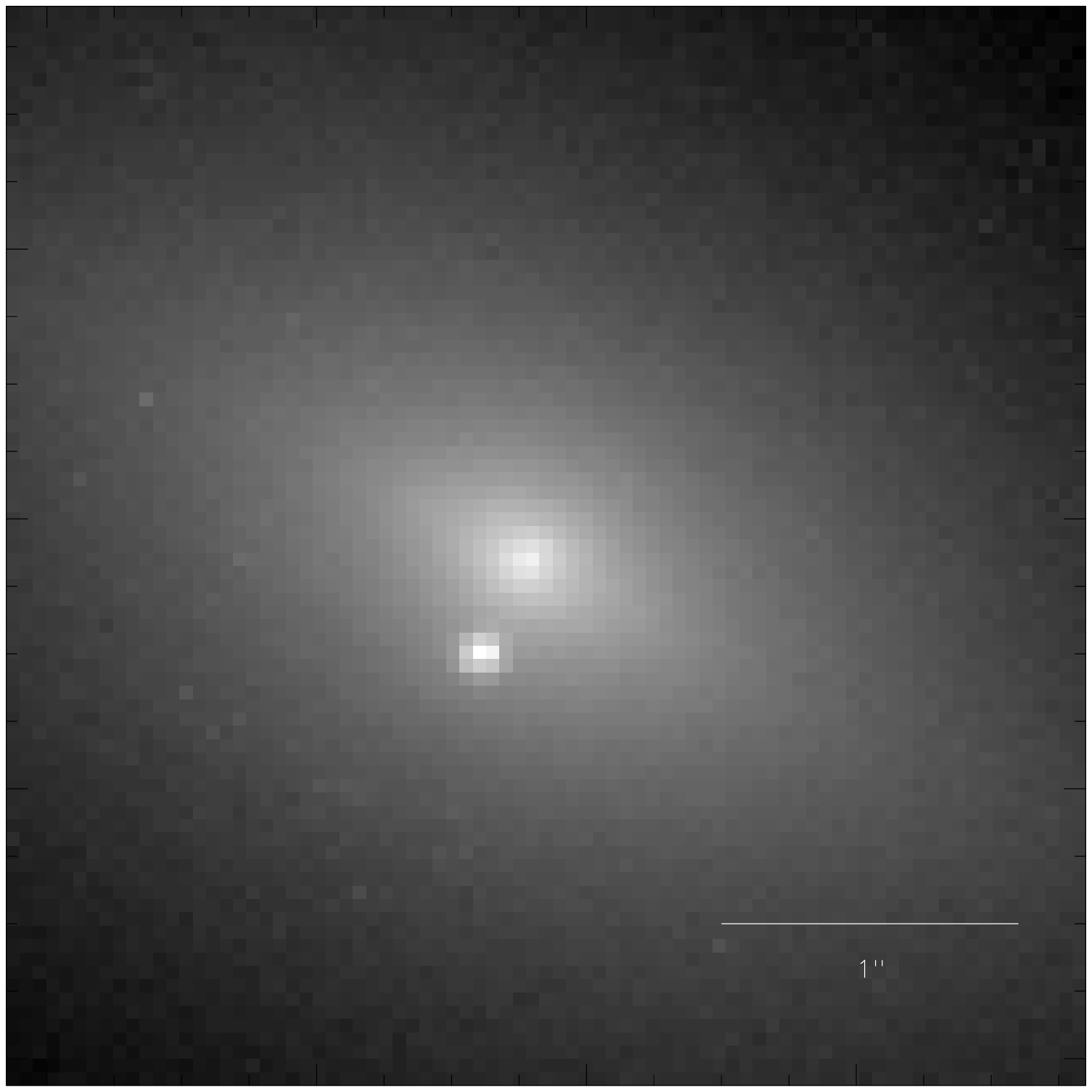} }}
\caption{\label{f:super} Zoom-in on the WFPC2 images of the nucleus of
   the NGC\,4128, in F702W (left) and F555W band (right). Both images
   are oriented as on Fig.~\ref{f:nuc_disk} (north is down and east to
   the right). Horizontal line is $1\arcsec$ long. The left image was
   taken almost two years before the right one. The offset of the peak
   next to the center on the right image is 0\farcs143 west and
   0\farcs318 north from the center of the galaxy. }
%\centering
\end{figure}

To determine the nature of the transient, it is necessary to
accurately do photometry on the observed light. The first required
step is to remove the contribution of the galaxy light. Since there
are no pre-transient observations of the galaxy in the two filters, it
was not possible to use the standard technique of subtracting a
suitable image of the galaxy taken with the same instrumental set up
\citep[e.g.][]{1986AJ.....92.1341F, 1994AJ....108.2226H}. To overcome
this problem a Multi-Gaussian Expansion (MGE) model of the galaxy was
constructed \citep[e.g.][]{1994A&A...285..723E}. The method and
software provided by \citet{2002MNRAS.333..400C} were used. The MGE
models of the galaxies were created using only the PC1 chip and
masking the region of the transient by a circular mask with radius of
5 pixel (Fig.~\ref{f:mge}). For both MGE models, the total number of
used Gaussians was increased until the minimum $\chi^{2}$ stopped
decreasing. This approach yielded the total number of 9 Gaussians and
RMS error of about 2\%.\looseness=-1

%%%%%%%%%%%Figure 7%%%%%%%%%%%%%%%%%%%%%%%%%%%%%%%%%%%%%%%%%%%%%%%%%%%%
\begin{figure}
{\hbox {\epsfxsize=0.48\columnwidth%\textwidth
\epsfbox{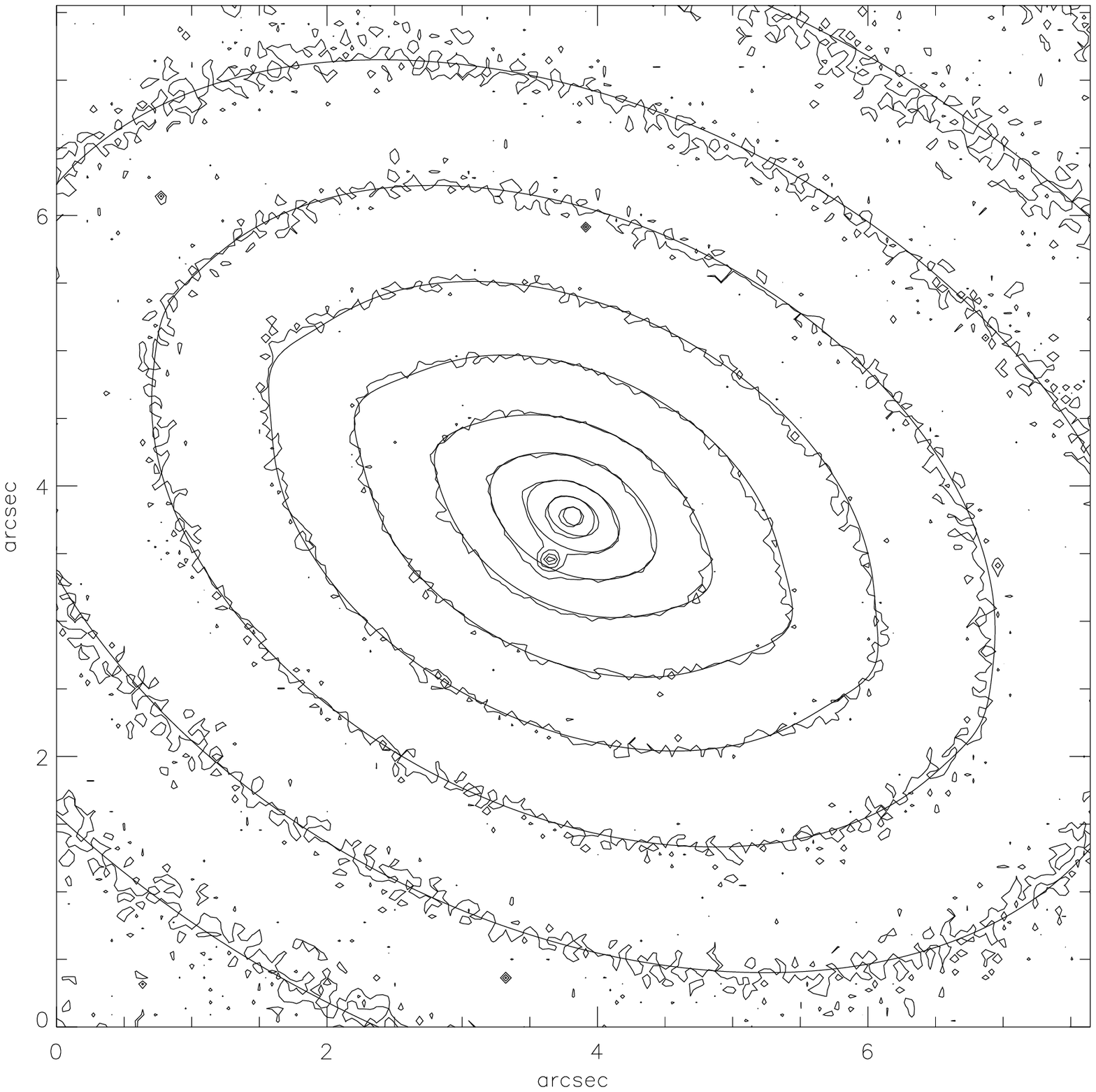}
\epsfxsize=0.48\columnwidth%\textwidth
\epsfbox{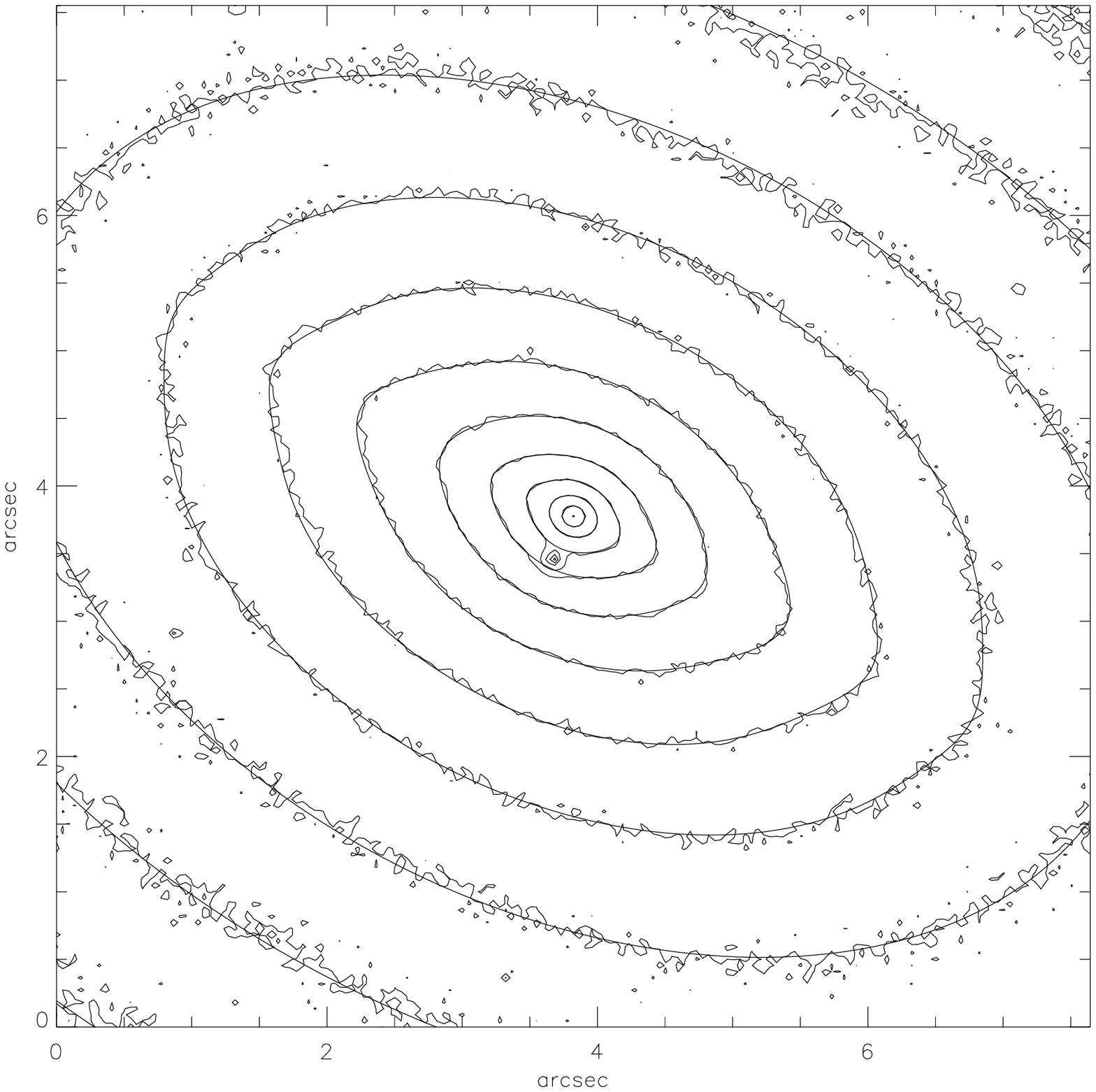} }}
\caption{\label{f:mge} Contour maps of WFPC2 F450W (left) and F555W
(right) images of NGC\'4128. The images show the inner
7\farcs5. Superposed on the two images are the contours of the MGE
surface brightness, convolved with the WFPC2 PSFs for the
corresponding filters. The transient was excluded from the fit, but it
is shown in the images for illustration.  }
\end{figure}

The obtained models were subtracted form the original
images. Figure~\ref{f:sup_prof} shows a horizontal cut through the
observed image at the position of the transient and the corresponding
MGE model. The light of the transient is clearly distinguishable from
the rest of the galaxy. Subtracting the MGE model of the galaxy from
the observations it is possible to recover the transient light
contribution. The residuals are symmetric and can be used to calculate
the magnitude of the transient. Photometric measurements were
conducted using the IRAF task PHOT. Magnitude zero points for filters
were obtained from \citet{2000PASP..112.1397D}. The values calculated
in an aperture of 0\farcs5 were corrected to infinite aperture by
adding 0.1 mag. The correction for geometrical distortion and the CTE
correction were not performed as their influence are of second order,
compared to the uncertainty introduced by subtracting the model galaxy
from the original image. The final transient parameters, namely its
position, apparent magnitude and absolute magnitude in Johnson-Cousins
system, are given in the Table~\ref{t:sup_obs}.

%%% Table B.1.%%%%%%%%%%%%%%%%%%%%%%%%%%%%%%%%%%%%%%%%%%%%%%%%%%%%%%%%%%%%%%%%
\begin{table*}
\begin{center}
 \caption[]{Summary of transient measurements.  }
  \label{t:sup_obs}
  \begin{tabular}{ccccccccc}
    \hline
    \hline
      filter & ra& dec & $\delta$ x& $\delta$ y& counts & m
      &$\sigma_{m}$ & M\\
      (1)&(2)&(3)&(4)&(5)&(6)&(7)&(8)&(9)\\
    \hline
    \hline
    F450W & 12:08:34.84 & 68:46:57.15 & 0.142 & 0.315 & 20940.09 & 19.06 & 0.05 &-13.7 \\
    F555W & 12:08:34.84 & 68:46:57.25 & 0.143 & 0.318 & 18099.38 & 19.17 & 0.05 &-13.6\\
    \hline
    \hline
 \end{tabular}
\end{center}

{(1): name of the filter; (2) and (3): position of the transient
measured in (h,m,s) and (deg,arcmin,arcsec) on WFPC2 (J2000); (4) and
(5): the offset in arcsec of the transient from the intensity peak of
the galaxy due west and north respectively; (6) total counts
inside aperture of 0\farcs5; (7) magnitude of the transient in
Johnson-Cousins system; (8) the estimated error on the apparent
magnitude; (9) absolute magnitude using the distance to NGC 4128 from
Table~\ref{t:prop}}

\end{table*}
%%%%%%%%%%%%%%%%%%%%%%%%%%%%%%%%%%%%%%%%%%%%%%%%%%%%%%%%%%%%%%%%%%%%%%%%%%%%%
\subsection{Discussion and conclusions}
\label{ss:dico}
Lacking any spectra of the detected object as well as a longer period
of observation, it is difficult to determine the object's true
nature. Since it is not visible on the F702W band image while it is
quite bright on the F450W and F555W band images (taken 22 months
later), it is most natural to conclude the object is a
transient. There are a few possibilities such as a solar system
object, a nova or a supernova.

%%%%%%%%%%%Figure 7%%%%%%%%%%%%%%%%%%%%%%%%%%%%%%%%%%%%%%%%%%%%%%%%%%%%
\begin{figure}
\centering
  {\hbox { \epsfxsize=0.5\columnwidth
      \epsfbox{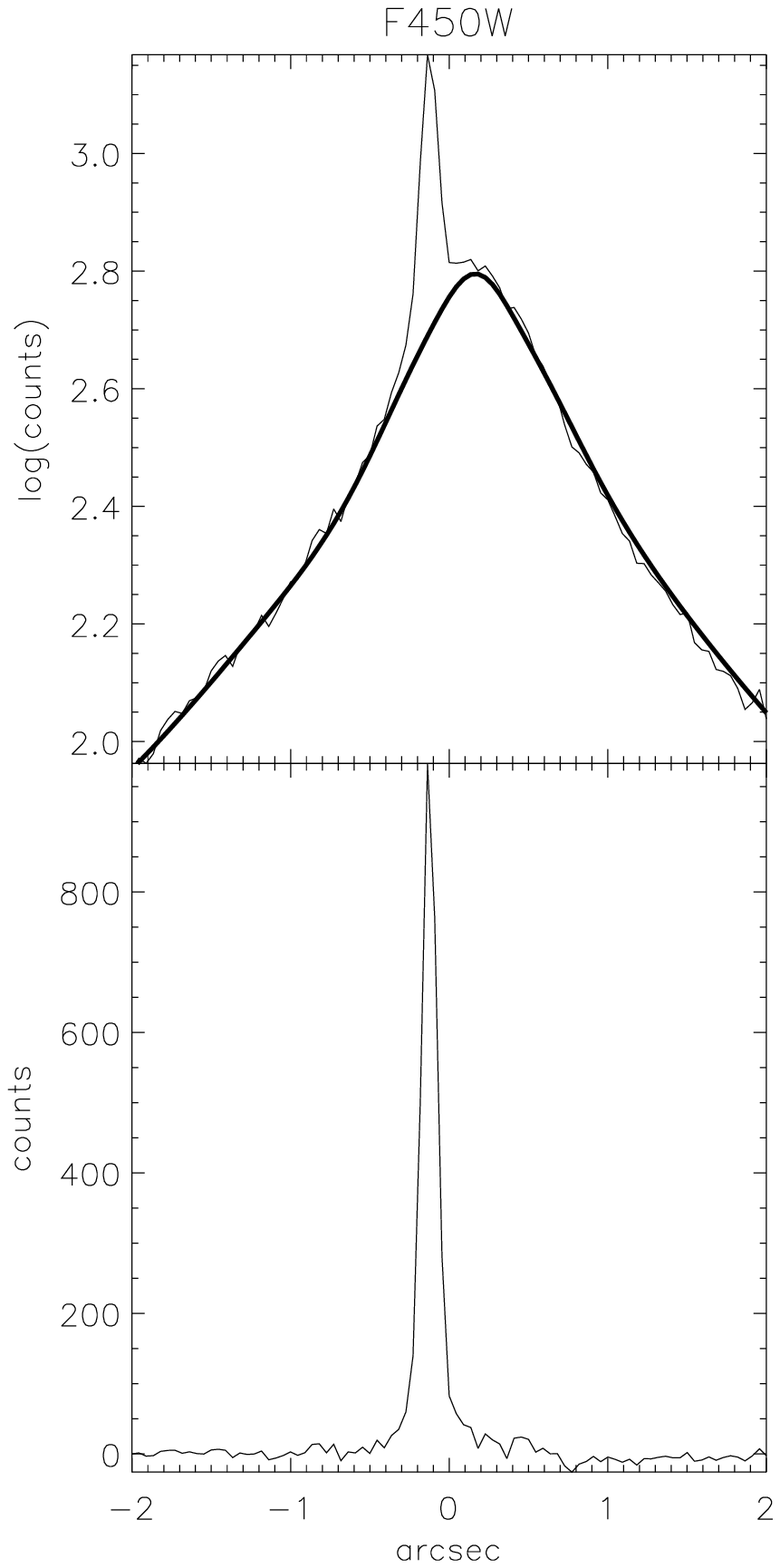}
      \epsfxsize=0.5\columnwidth
      \epsfbox{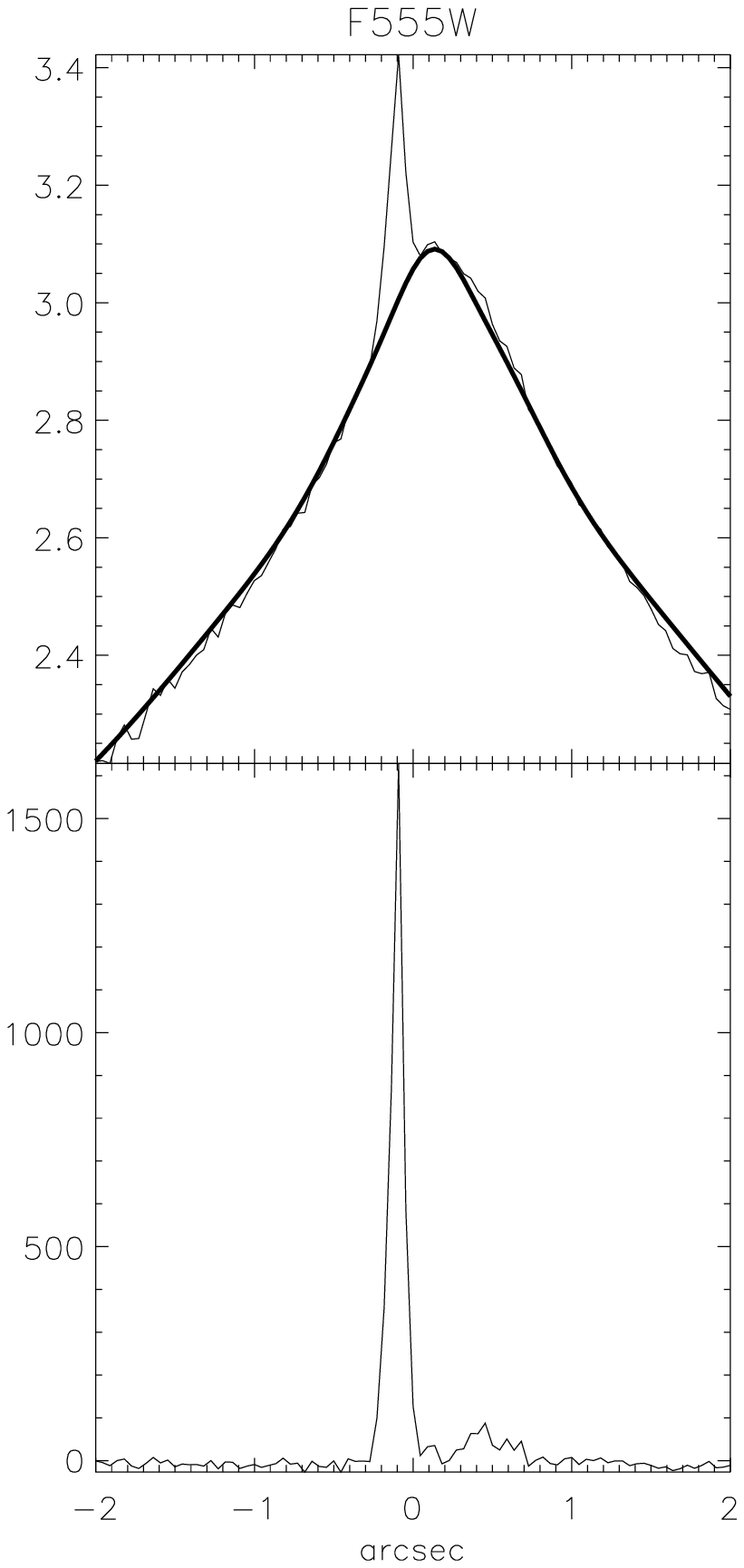} }}
  \caption{\label{f:sup_prof}Horizontal cuts through the galaxy on the
position of the SN. The thick line is the MGE model and the thin line
is the observed light. The lower plots show residuals obtained
subtracting the models from the observations. The transient is clearly
separable from the rest of the galaxy by a MGE model.}  \centering
\end{figure}

The duration of all four independent WFPC2 exposures is about 40
minutes. A solar system object would show a noticeable movement
between the different exposures, except if moving directly toward an
observer. If the object is at the distance of the Kuiper belt it would
move approximately a few arc seconds during the observations. However,
the relative position of the object (the distance between the object
and the center of the galaxy) changes by less than one pixel ($<$
0\farcs0455), the movement being less than measurement error. It seems
reasonable to conclude the object is of extra-Solar system origin.

Novae are known to have a maximum absolute magnitude less bright then
-9 mag \citep{1985ApJ...292...90C}. With the measured absolute
magnitude of -13.6 (V band), this is ruled out. Similarly, because it
is very faint, the observed transient is also not likely a nova in our
galaxy (a compilation of novae light curves is given in
\citet{1987A&AS...70..125V}). The observed magnitude leaves the
possibility that the object in NGC\,4128 is a supernova. The type of
supernova is defined according to its spectrum, and here again it is
hard to establish anything specific. However, the host galaxy is an S0
galaxy, and since supernovae of type II do not occur in early-type
galaxies \citep[Table 3. in][]{1997A&A...322..431C}, the transient is
probably a supernova of Type Ia. Comparing the measured absolute
magnitude to absolute magnitudes of supernova 1994D or 1992A
\citep{1997A&A...328..203C}, which were also S0 galaxies, suggests
that the supernova was observed about 200 days after the
explosion. The $B-V$ color of the supernova is -0.11, which is bluer
then the expected color for a type Ia SN at B maximum. On the other
had, it is consistent with the $B-V$ light curve at the later time
(around 200 days) of the above mentioned reference supernovae
\citep{2001MNRAS.321..254S}. \looseness=-2

Private communication with the supernova-survey groups (D.~Green from
IAU Circulars and W.~Li from LOTOSS survey) did not confirm the
existence of the supernova. They checked images from January to June
2001, and after galaxy subtraction there was no signature of any
transient object. However, the proximity to the nucleus of NGC\,4128
is a possible reason why ground-based automated search program might
have overlooked the supernova. \looseness=-2

If the source of the light is a Type Ia supernova, it will be the
first supernova discovered in NGC\,4128
\citep{1999A&AS..139..531B}\footnote{
\texttt{http://web.pd.astro.it/supern/}}. Unfortunately, without other
high resolution available data it is impossible to determine the true
nature of the object. Also, the estimate of the timing of the
supernova in NGC\,4128 is highly uncertain, since it depends on
precise classification and calibration. Type Ia supernovae are known
to have different absolute magnitude light curves
\citep[e.g.][]{1997A&A...328..203C}. If the transient in the NGC\,4128
is not a supernova of type Ia 200 days after explosion then its nature
remains unknown (a flaring second black hole, an asteroid on the
trajectory to Earth).

\end{document}